\documentclass[final,3p,times]{elsarticle}
\usepackage{amsfonts}
\usepackage{mathrsfs}
\usepackage{amsmath}
\usepackage{xcolor}
\usepackage{amssymb}
\usepackage{bm}
\usepackage{feynmp}
\usepackage{extarrows}
\usepackage{slashed}
\usepackage{graphicx}
\usepackage{subfigure}
\usepackage[linktocpage=true,bookmarks=true,bookmarksnumbered=true,breaklinks=true,pdfpagemode=Fullscreen,pdfstartview=FitBH]{hyperref}

\newcommand{\fr}[2]{\mbox{$\frac{\,{#1}\,}{#2}$}}

\renewcommand{\rm}{\mathrm}
\def\bge{\begin{equation}}
\def\ede{\end{equation}}
\def\bga{\begin{aligned}}
\def\eda{\end{aligned}}
\newcommand{\beq}{\begin{equation}}
\newcommand{\eeq}{\end{equation}}
\newcommand{\bq}{\begin{equation}}
\newcommand{\eq}{\end{equation}}
\newcommand{\ba}{\begin{array}}
\newcommand{\ea}{\end{array}}
\newcommand{\beqa}{\begin{eqnarray}}
\newcommand{\eeqa}{\end{eqnarray}}
\newcommand{\beqs}{\begin{subequations}}
\newcommand{\eeqs}{\end{subequations}}

\def\dis{\displaystyle}
\def\ff{\frac}
\def\({\left(}
\def\){\right)}
\def\[{\left[}
\def\]{\right]}

\def\End{\end{document}}

\def\leqq{\leqslant}

\def\hf{\frac{1}{2}}

\def\ot{\otimes}

\def\al{\alpha}

\def\ga{\gamma}

\def\ga{g_0^{}}
\def\gaa{g_0^2}
\def\gb{g_1^{}}
\def\gbb{g_1^2}
\def\gc{g_2^{}}
\def\gcc{g_2^2}

\def\T{\mathcal{T}}
\def\B{\mathcal{B}}

\def\mb{\mathbf}

\def\tr{\mathrm{\,tr\,}}

\def\to{\rightarrow}

\def\wt{\widetilde}
\def\PA{\Phi_1^{}}
\def\PB{\Phi_2^{}}

\def\Shat{\widehat{S}}
\def\That{\widehat{T}}

\def\bbbar{b\bar{b}}
\def\tautaub{\tau\bar{\tau}}
\def\gaga{\gamma\gamma}

\def\T{{\mathcal{T}}}

\newcommand{\mT}{\mathcal{T}}
\newcommand{\mB}{\mathcal{B}}
\newcommand{\mL}{\mathcal{L}}
\newcommand{\mO}{\mathcal{O}}
\newcommand{\mR}{\mathcal{R}}
\newcommand{\ca}{c_{\alpha}^{}}
\newcommand{\sa}{s_{\alpha}^{}}

\newcommand{\cw}{c_{W}^{}}
\newcommand{\sw}{s_{W}^{}}
\newcommand{\cww}{c_{W}^2}
\newcommand{\sww}{s_{W}^2}
\newcommand{\TeV}{\textrm{TeV}}
\newcommand{\GeV}{\textrm{GeV}}

\begin{document}

\begin{frontmatter}

\setcounter{footnote}{1}
\renewcommand{\thefootnote}{\fnsymbol{footnote}}

\title{{\bf LHC Higgs Signatures from Topflavor Seesaw Mechanism}}

\author{{\sc Xu-Feng Wang} ~and~ {\sc Chun Du}}

\address{Institute of Modern Physics and Center for High Energy Physics,
                Tsinghua University, Beijing 100084, China}

\author{{\sc Hong-Jian He}\,\footnote{Corresponding email: hjhe@tsinghua.edu.cn}}

\address{Institute of Modern Physics and Center for High Energy Physics,
                Tsinghua University, Beijing 100084, China\\
Center for High Energy Physics, Peking University, Beijing 100871, China\\
Kavli Institute for Theoretical Physics China, CAS, Beijing 100190, China
}

\begin{abstract}
We study LHC Higgs signatures from topflavor seesaw realization of
electroweak symmetry breaking with a minimal gauge extension
$\,SU(2)\ot SU(2) \ot U(1)\,$.\, This elegant renormalizable construction
singles out top quark sector (instead of all other light fermions)
to join the new $SU(2)$ gauge force.  It predicts extra vector-like
spectator quarks $\,(\T,\,\B)$\,,\, new gauge bosons \,($W',\,Z'$)\,,\,
and a pair of neutral Higgs bosons \,($h,\,H$)\,.\,
We demonstrate that for the lighter Higgs boson $\,h\,$ of mass 125\,GeV,
this model predicts modified Higgs signal rates in
$\,h\to\gamma\gamma,\,WW^*,\,ZZ^*$\, channels via gluon fusions,
in $\,h\to \tautaub\,$ mode via vector boson fusions, and
in $\,h\to \bbbar\,$ mode via gauge boson associate productions.
We perform a global fit for our theory by including both direct search data
(LHC and Tevatron) and indirect precision constraints.
We further analyze the LHC discovery potential
for detecting the heavier Higgs state $\,H$\,.
\\[2mm]
\\[2mm]
\hfill
Phys.\ Lett.\ B\,(published version), [\,arXiv:1304.2257\,]
\end{abstract}

\end{frontmatter}





\renewcommand{\thefootnote}{\arabic{footnote}}

\section{Introduction}
\vspace*{2mm}

 With the exciting LHC discovery of a Higgs-like new boson of mass around
 125\,GeV \cite{LHC2012}\cite{LHC2013-3}\cite{LHC2013-3b},
 we study the prediction of Higgs signals in the topflavor seesaw model
 proposed in \cite{He:1999vp}.
 This model is strongly motivated by the heavy top quark with large mass
 $\,m_t^{}\simeq 173.18\pm 0.94$\,GeV\, \cite{mt},
 which stands out at the weak scale together with weak gauge bosons $(W,\,Z)$.\,
 All other standard model (SM) fermions have masses no more than $\,\mO(\GeV)$\,.\,
 Hence, it is truly attractive to expect that the top sector may invoke certain
 new gauge dynamics at the weak scale, but all other light fermions (including
 the third family tau lepton) do not.
 It was realized \cite{He:1999vp} that such a construction enforces the
 introduction of extra spectator quarks for gauge anomaly cancellation,
 and thus unavoidably leads to seesaw mechanism for top mass generation.
 This elegant renormalizable construction was called the topflavor seesaw \cite{He:1999vp},
 where the top sector joins an extra new $SU(2)$ or $U(1)$ gauge force.
 It differs from the traditional topcolor seesaw models \cite{He:2001fz}
 involving strong $SU(3)$ topcolor gauge group with singlet heavy quarks,
 as well as the early non-universality model with an extra $SU(2)$
 for the whole third family \cite{NU}. It also differs from the
 ununified model\,\cite{UU} which has quarks and leptons couple
 to two separate $SU(2)$'s.

 In this Letter, we study
 an explicit realization of topflavor seesaw via gauge group
 $\,{\cal G}= SU(3)_c\otimes SU(2)_t^{}\otimes SU(2)_f^{} \otimes U(1)_y^{}\,$,
 (called type-I \cite{He:1999vp}). It invokes two Higgs doublets
 $\Phi_1^{}$ and $\Phi_2^{}$ to spontaneously break $\,{\cal G}\,$ down to
 the residual symmetry $\,SU(3)_c\otimes U(1)_{\text{em}}\,$.\,
 In consequence, two neutral physical Higgs boson $h^0$ and $H^0$ are
 predicted, in addition to the weak gauge bosons $(W,\,Z)$ and $(W',\,Z')$.
 Ref.\,\cite{He:1999vp} focused on the construction of
 Yukawa sector for topflavor seesaw
 and the electroweak precision constraints on the spectator quarks.
 In this work, we will systematically study the Higgs sector of this model
 and derive new predictions for the $h^0$ and $H^0$ Higgs signatures at the LHC.
 We also note that a renormalizable flavor universal construction of the
 electroweak gauge group $SU(2)_0^{}\otimes SU(2)_1^{}\otimes U(1)_2^{}$
 (the 221 model) was recently studied in \cite{Abe:2012fb} together
 with its Higgs phenomenology at the LHC (which serves as an ultraviolet
 completion of the conventional three-site model \cite{Chivukula:2006cg}).
 Our current topflavor seesaw model
 shares similarity with the 221 model \cite{Abe:2012fb}
 in the gauge group and Higgs sector, because their structures of spontaneous
 symmetry breaking both belong to the three-site linear moose representation.
 However, the topflavor seesaw differs from the 221 model in several essential ways:
 (i) the topflavor seesaw embeds fully different fermion assignments
     under the gauge group, and singles out the mass-generation of top sector from
     all other light SM fermions;
 (ii) it embeds only a pair of the spectator quarks $(\T,\,\B)$, associated
       with top sector; the $(\T,\,\B)$ are vector-like under the diagonal subgroup
      of $\,SU(2)_t^{}\otimes SU(2)_f^{}\,$ after spontaneous symmetry breaking,
      but not under one of the parent $SU(2)$'s;
 (iii) the ranges of expansion parameters in terms of the gauge coupling ratio and
      the ratio of Higgs vacuum expectation values (VEVs)
      fully differ from those of the 221 model.
 In consequence, our current study will present fully different new
 predictions for the LHC signals of Higgs bosons, heavy gauge bosons and
 vector-like fermions.

 This Letter is organized as follows. In Sec.\,2, we analyze the gauge and Higgs sectors
 of the topflavor seesaw model.  We also derive the direct and indirect bounds
 on the new gauge bosons $(W',\,Z')$. In Sec.\,3, we study the LHC signals of the
 lighter Higgs boson $\,h^0\,$ of mass $125$\,GeV. Sec.\,4 is devoted to the analysis
 of LHC potential of detecting the heavier Higgs state $H^0$. Finally, we conclude in
 Sec.\,5.


\vspace*{2mm}
\section{Topflavor Seesaw: Structure, Parameter Space and Constraints}

\vspace*{2mm}
\noindent
{\bf 2.1. Structure of the Model: Gauge, Higgs and Yukawa Sectors}
\vspace*{3mm}

As mentioned above, the large top mass $\,m_t^{}\simeq v/\!\sqrt{2}\simeq 173\,$GeV\,
stands out of all SM fermions, suggesting that the top sector is special and
may invoke a new gauge force, but all other light SM fermions (including tau lepton)
do not. It was found\,\cite{He:1999vp} that anomaly cancellation
enforces the introduction of spectator quarks $\,S=(\T,\,\B)^T\,$
and generically leads to the seesaw mechanism for top mass generation.
In the present work, we will focus on
the topflavor seesaw gauge group of type-I \cite{He:1999vp},
$\,{\cal G}= SU(3)_c\otimes SU(2)_t^{}\otimes SU(2)_f^{} \otimes U(1)_y^{}\,$,\,
where only the top-sector enjoys the extra $\,SU(2)_t^{}$\, gauge forces
which is stronger than the ordinary $SU(2)_f^{}$ (associated
with all other light fermions). Hence, the structure of topflavor seesaw
is completely fixed, which is anomaly-free and renormalizable.
This is summarized in Table\,\ref{tab:1}, where
we only show the assignments for the third family fermions and
Higgs sector. All the first two families of fermions are charged under
$\,SU(3)_c^{}\ot SU(2)_f^{}\ot U(1)_y^{}$\, in the same way as in the SM.

\begin{table}[h]
\begin{center}
\begin{tabular}{c||cccc}
\hline\hline
&&&&\\[-3mm]
~~Fields~~ & ~$SU(3)_c^{}$~ & ~$SU(2)_t^{}$~ & ~$SU(2)_f^{}$~ & ~$U(1)_y^{}$~
\\[0.15cm]
\hline\hline
&&&&\\[-3mm]
$Q_{3L}^{}$   & \bf{3}        & \bf{2}        & \bf{1}  & ~~$\ff{1}{6}$~~
\\[1.5mm]
~$(t_R^{},b_R^{})$~  & \bf{3}  & \bf{1}  & \bf{1}   & ~~$\(\ff{2}{3},-\ff{1}{3}\)$~~
\\[1.5mm]
$S_L^{}$      & \bf{3}        & \bf{1}        & \bf{2}        & $\ff{1}{6}$
\\[1.5mm]
$S_R^{}$      & \bf{3}        & \bf{2}        & \bf{1}        & $\ff{1}{6}$
\\[1.5mm]
$L_3^{}$      & \bf{1}        & \bf{1}        & \bf{2}        & $-\hf$~\,
\\[1.5mm]
$\tau_R^{}$   & \bf{1}        & \bf{1}        & \bf{1}        & $-1$~\,
\\[1.5mm]
\hline
&&&&\\[-3mm]
$\Phi_{1}^{}$     & \bf{1}        & \bf{2}        & \bf{2}        & $0$
\\[1.5mm]
$\Phi_{2}^{}$     & \bf{1}        & \bf{1}        & \bf{2}        & $\hf$
\\[1mm]
\hline\hline
\end{tabular}
\end{center}
\vspace*{-3mm}
\caption{Anomaly-free assignments of the third family fermions and the Higgs sector
         in (type-I) topflavor seesaw, where $\,Q_{3L}^{}=(t,\,b)_L^{T}$,\,
         $\,L_{3}^{}=(\nu_{\tau}^{},\,\tau)_L^{T}$,\, $\,S=(\T,\,\B)^{T}$,\,
         and the hypercharge is defined via $\,Q=I_3^{}+Y$\,.\,}
\label{tab:1}
\end{table}

The electroweak part of the gauge group,
$\,SU(2)_t^{}\otimes SU(2)_f^{} \otimes U(1)_y^{}\,$,\,
forms a three-site linear moose, from left to right.
We denote the corresponding three gauge couplings as
\,$(\ga,\,\gb,\,\gc)$\,,\, and will consider the parameter
space with $\,\gaa\gg\gbb > \gcc\,$.\,
(This differs from the 221 model\,\cite{Abe:2012fb}
and 3-site model\,\cite{Chivukula:2006cg}
which define the parameter region $\,\gbb\gg\gaa > \gcc\,$ instead.)
For the Higgs sector, the two Higgs doublets $\PA$ and $\PB$ transform under
$\,SU(2)_t^{}\otimes SU(2)_f^{} \otimes U(1)_y^{}\,$
as $(\mb{2},\mb{2},0)$ and $(\mb{1},\mb{2},\hf)$, respectively.
Thus, we can write them in the self-dual quartet form,
\beqa
\PA \,=\, u+h_{1}^{}+i\vec{\tau}\cdot\vec{\pi}_{1}^{}\,,
~~~~~~
\PB \,=\, v+h_{2}^{}+i\vec{\tau}\cdot\vec{\pi}_{2}^{}\,,
\eeqa
which develop nonzero VEVs, \,$u\gg v$,\, from the Higgs potential.
This breaks the gauge symmetry as follows,
\begin{subequations}
\begin{eqnarray}
\label{eq:break1}
SU(2)_{t}^{}\otimes SU(2)_{f}^{}
&~\xrightarrow{~\langle\PA\rangle\,=\,u~}~
& SU(2)_{L}^{}\,,
\\[1mm]
\label{eq:break2}
SU(2)_{L}^{} \otimes U(1)_y^{}
&~\xrightarrow{~\langle\PB\rangle\,=\,v~}~
& U(1)_{\textrm{em}}^{} \,.
\end{eqnarray}
\end{subequations}
In consequence, it results in the coupling relation,
$\,g_0^{-2} + g_1^{-2} + g_2^{-2} = e^{-2}\,$.\,
Then, we present the Lagrangian of the gauge and Higgs sectors,
\begin{eqnarray}
\mL &\!=\!\!&
-\frac{1}{4} \sum_{a = 1}^{3} V^{a}_{0 \mu \nu} V^{a \mu \nu}_{0}
-\frac{1}{4} \sum_{a = 1}^{3} V^{a}_{1 \mu \nu} V^{a \mu \nu}_{1}
-\frac{1}{4} V_{2 \mu \nu}^{} V^{\mu \nu}_{2}
\,+\, \frac{1}{4}\sum_{j=1,2} \!\tr\!\Big[(D_\mu \Phi_j)^\dagger (D^\mu \Phi_j)\Big]
    -V(\Phi_1, \Phi_2)
\,,~~~~
\label{eq:L}
\end{eqnarray}
where the gauge field strengths
$V^{a \mu \nu}_{0}$, $V^{a \mu \nu}_{1}$, and $V^{\mu \nu}_{2}$
are associated with $SU(2)_{t}^{}$, $SU(2)_{f}^{}$ and $U(1)_y^{}$,\,
respectively. The covariant derivatives for Higgs fields are given
by,
$\,D_{\mu}^{} \PA
= \partial_{\mu}^{} \PA + ig_0^{}\frac{\tau^{a}}{2}V_{0\mu}^{a}\PA
  - ig_1^{} \PA\frac{\tau^{a}}{2} V_{1\mu}^{a}$ \,
and
$\,D_{\mu}^{} \PB
= \partial_{\mu}^{} \PB + ig_1^{}\frac{\tau^{a}}{2}V_{1\mu}^{a}\PB
  - ig_2^{}\PB\frac{\tau^{3}}{2} V_{2\mu} $\,.\,

Then, we can readily derive the mass-matrices for the charged and neutral
gauge bosons, as follows,
\beqa
\label{eq:WZ-MassM}
\mathbb{M}_W^2 =\,
\frac{\,g^{2}_{0}u^2}{4}
\left\lgroup\!\!\!
\begin{array}{cc}
1 & -x
\\[1.5mm]
-x & x^{2}(1\!+\!y^{2})
\end{array}
\!\!\!\right\rgroup\!,
&~~~&
\mathbb{M}_N^2 =\,
\frac{\,g^{2}_{0}u^2}{4}
\left\lgroup\!\!\!
\begin{array}{ccc}
1 & -x & 0
\\[1.5mm]
-x & x^{2}(1\!+\!y^{2}) & -x^{2}y^{2}t
\\[1.5mm]
0 & -x^{2}y^{2}t & x^{2}y^{2}t^{2}
\end{array}
\!\!\!\right\rgroup\!,
\eeqa
where we have defined the ratios, $\,y \equiv v/u\,$,\,
      $\,x \equiv \gb/\ga$\,,\,
and   $\,t \equiv \gc/\gb$\,.\,
Our construction sets the parameter space,  $\,x^2,y^2 \ll 1\,$.\,
Thus, we can expand the masses and couplings in power series of
$\,x\,$ and $\,y\,$.\,
With these we infer the mass-eigenvalues of charged and neutral weak
bosons, $(W,\,W')$ and $(Z,\,Z')$, from diagonalizing (\ref{eq:WZ-MassM}),
%
\beq
\label{eq:VV'-mass}
\ba{ll}
M_{W}^{} ~=~ \dis
\frac{ev}{\,2\sw\,}\(1-\fr{1}{2}x^{4}y^{2}) + \mO(x^{6}\) \,,
~~~&~~~
M_{W'}^{} ~=~ \dis
\frac{ev}{\,2\sw\,xy\,}\(1+x^{2}+\fr{1}{2}x^{4}y^{2}\) + \mO(x^{5})\,,
\\[4mm]
M_{Z}^{} ~=~ \dis
\frac{ev}{\,2\sw\cw\,}\(1-\fr{1}{2}x^{4}y^{2}\) + \mO(x^{6})\,,
~~~&~~~
M_{Z'}^{} ~=~ \dis
\frac{ev}{\,2\sw\,xy\,}\(1+x^{2}+\fr{1}{2}x^{4}y^{2}\) + \mO(x^{5})\,,
\ea
\eeq
%
where we have used notations $\,(\sw ,\, \cw) \equiv (\sin\theta_W^{},\,\cos\theta_W^{})\,$.\,
In the above, we have also defined, $\,g^{-2}\equiv g_0^{-2}+g_1^{-2}\,$ and
$\,\tan\theta_W^{}\equiv \gc/g\,$, which lead to $\,e=g\sw\,$.\,
Eq.\,(\ref{eq:VV'-mass}) gives the mass ratios,
$\,M_{W'}^{}/M_W^{} \simeq M_{Z'}^{}/M_W^{}\simeq (1+x^2)/(xy)\,$.

For the Higgs sector, we can write down the general gauge-invariant and
CP-conserving Higgs potential of $\,\PA\,$ and $\,\PB\,$ as follows,
\begin{eqnarray}
 V(\Phi_1^{}, \Phi_2^{}) \,=\,
 \frac{1}{2}\lambda_1^{}
 \!\left[ \frac{1}{4}\tr(\Phi_1^\dagger \Phi_1^{}) -\dfrac{u^2}{2} \right]^2\!
  +\frac{1}{2}\lambda_2^{}\!
   \left[ \frac{1}{4}\tr(\Phi_2^\dagger \Phi_2^{}) -\dfrac{v^2}{2} \right]^2\!
+\lambda_{12}^{}
  \left[ \frac{1}{4}\tr(\Phi_1^\dagger \Phi_1^{}) -\dfrac{u^2}{2} \right]\!
  \left[ \frac{1}{4}\tr(\Phi_2^\dagger \Phi_2^{}) -\dfrac{v^2}{2} \right] .
\hspace*{3mm}
\label{eq:V}
\end{eqnarray}
After spontaneous symmetry breaking, the six gauge bosons
$(W,\,Z)$ and $(W',\,Z')$ absorb the corresponding would-be
Goldstone bosons $(\pi_1^a,\,\pi_2^a)$ and acquire masses via
Higgs mechanism \cite{HM}.
We derive the mass-eigenvalues of the two remaining physical Higgs bosons
$h^0$ and $H^0$, which are connected to the weak eigenstates
$(h_1^{},h_2^{})$ via a $2\times 2$ orthogonal rotation with mixing
angle $\,\alpha$\,.\, Thus, we arrive at,
\begin{subequations}
\label{eq:HiggsM-alpha}
\begin{eqnarray}
\label{eq:HiggsMass}
M_{h,H}^2 &\!\!=\!\!&
\frac{\,vu\,}{2}\left[\(\lambda_1^{}y^{-1} \!+\lambda_2^{}y\) \mp \!
\sqrt{(\lambda_1^{}y^{-1}\! - \lambda_2^{}y)^2 + 4\lambda_{12}^2\,}\,\right],
\\[1.5mm]
\tan 2\alpha &\!\!=\!\!&
2\lambda_{12}^{}\, / \(\lambda_2^{}y - \lambda_1^{}y^{-1}\) \,,
\label{eq:mixing}
\end{eqnarray}
\end{subequations}
where the range of $\,\alpha\,$ is chosen as, $\,\alpha\in [0,\pi)\,$.\,
We note that the Higgs potential (\ref{eq:V}) has five parameters in total,
two Higgs VEVs $\,(v,\,u)$\, and three self-couplings
$\,(\lambda_1^{},\,\lambda_2^{},\,\lambda_{12}^{})\,$.\,
The VEV $\,v\simeq 246\,$GeV\, will be fixed by the Fermi constant as in (\ref{eq:GF-v})
and $\,u\,$ can be converted to the ratio $\,y\equiv v/u\,$.\,
Under $\,M_h^{}=125\,$GeV,\, the three Higgs self-couplings will be fully fixed
by inputting the heavier Higgs mass $M_H^{}$ and mixing angle $\,\al\,$.
We will further constrain the three independent parameters $\,(y,\,\al,\,M_H^{})\,$
from the global fit in Sec.\,3.

The topflavor seesaw mechanism is realized in the Yukawa sector.
According to the assignments of Table\,\ref{tab:1}, we have the following
Yukawa interactions for the top sector \cite{He:1999vp},
\beqa
\label{eq:L-seesaw}
\mL_{Y}^{t} &\!\!=\!\!&\!
-\frac{\,y_{s}^{}}{\sqrt{2}\,}\overline{S_{L}^{}}\PA S_{R}^{}
+y_{st}^{}\overline{S_{L}^{}}\wt{\Phi}_2't_{R}^{}
+y_{sb}^{}\overline{S_{L}^{}}\Phi_2'b_{R}
-\kappa\,\overline{Q^{}_{3L}}S_{R}^{} + \text{h.c.}\,,
\eeqa
where we have reexpressed the second Higgs field $\PB$ in terms of the usual doublet form,
$\,\Phi_2' =\PB (0,\,\frac{1}{\sqrt 2})^T
           = (i\pi_2^{+},\,\frac{1}{\sqrt 2}(v+h_2^{}-i\pi_2^{0}))^{T}$.\,
From Eq.\,(\ref{eq:L-seesaw}), we deduce
the seesaw mass matrices for top and bottom quarks,
\begin{equation}
-\(\overline{t_L^{}},\,\overline{\mT_L^{}}\)\!
\left\lgroup\!\!\!
\begin{array}{cc} 0  & \kappa \\[1mm]
                    -m_{st} & M_{S}^{}
\end{array} \!\!\!\right\rgroup \!
\left(\!\! \begin{array}{c} t_R \\[1mm] \mT_R \end{array} \!\!\right)
- \(\overline{b_L^{}},\, \overline{\mB_L^{}}\) \!
\left\lgroup\!\!\!  \begin{array}{cc} 0       & \kappa \\[1mm]
                                    -m_{sb}^{} & M_{S}^{}
  \end{array}  \!\!\!\right\rgroup\!
\left(\!\! \begin{array}{c} b_R\\[1mm] \mB_R \end{array} \!\!\right)
+ \text{h.c.}\,,
\label{eq:SeesawMass}
\end{equation}
where $\,M_S^{} = y_s^{}u/\sqrt{2}$\,,\, $m_{st}^{}=y_{st}^{}v/\sqrt{2}$\,,\, and
      $\,m_{sb}^{}=y_{sb}^{}v/\sqrt{2}$\,.\,
The $\,\kappa\,$ mass-term in (\ref{eq:L-seesaw}) is gauge-invariant,
and is expected to be around $\,\mO(M_S^{})$\,.\,
Diagonalizing the seesaw mass-matrices in Eq.\,(\ref{eq:SeesawMass}), we have the
following mass-eigenvalues for top, bottom, and their spectators,
\beqs
\begin{eqnarray}
m_{t(b)}^{} &\!\!=\!\!&
\frac{m_{st(sb)}^{}\,\kappa}{M_S^{}\!\sqrt{1\!+\!r\,}\,}
     \!\left[ 1-\frac{\,m_{st(sb)}^2/M_S^2\,}{2(1\!+\!r)^2}
      +\mO\left(\frac{m_{t(b)}^4}{M_S^4}\right) \right]\!,
\\
M_{\mT (\mB )}^{} &\!\!=\!\!&
M_S^{}\!\sqrt{1+r\,}\,\left[
1 + \frac{z_{t(b)}^2}{2(1\!+\!r)} +
\frac{4r\!+\!3}{8(1\!+\!r)^2} z_{t(b)}^4 + \mO(z_{t(b)}^6)
\right]\!,
\label{eq:mtbTB}
\end{eqnarray}
\eeqs
where we have defined the ratios
$\,\sqrt{r} \equiv \kappa/M_{S}^{} = \mO(1)$\, and
$\,z_{t(b)}^{} \equiv m_{t(b)}^{}/\kappa$\, with
$\,z_{b}^{}\ll z_{t}^{}\ll 1$\,.\,
Note that the heavy quarks \,$(\mT,\mB)$\, are highly degenerate
because their mass-splitting
$\,M_{\mT}^{} - M_{\mB}^{} \simeq
   m_t^{}z_t^{}/\!\sqrt{4r(1\!+\!r)\,} \ll m_t^{}\,$.\,
The diagonalization of seesaw mass-matrices (\ref{eq:SeesawMass})
is realized by the $2\times 2$ bi-unitary rotations,
$\,{U_L^{j\dag}}\mathbb{M}^jU_R^j = \mathbb{M}^j_{\text{diag}}\,$,\, where
the index $\,j =t,b\,$ denotes the up-type and down-type transformations.
The corresponding rotation angles for the seesaw diagonalizations are,
\beqs
\label{eq:ss-angle}
\begin{eqnarray}
\sin\theta_R^j &\!\!=\!\!&
-\frac{z_j}{\sqrt{1\!+\!r}\,}\!
\left[1+\frac{r}{\,1\!+\!r\,}z_j^2\right] + O(z_j^5)\,,
\label{eq:ss-angleR}
\\[1.5mm]
\sin\theta_L^j &\!\!=\!\!&
\sqrt{\frac{r}{\,1\!+\!r\,}\,} \left[1-\frac{z_j^2}{\,1\!+\!r\,}
         -\frac{3r}{\,2(1\!+\!r)^2\,}z_j^4 + O(z_j^6)\right].
\label{eq:ss-angleL}
\end{eqnarray}
\eeqs
We note that the right-handed rotation is suppressed by
$\,z_{t(b)}^{} \equiv m_{t(b)}^{}/\kappa \ll 1\,$,\,
and especially $\,\theta_R^b$\, is negligible since
$\,z_b^{}/z_t^{}=m_b^{}/m_t^{}\approx 1/40\,$.\,

\vspace*{5mm}
\noindent
{\bf 2.2. Parameter Space: Indirect and Direct Constraints}
\vspace*{3mm}

For analysis of the indirect precision constraints, we will
follow the formalism of \cite{Barb}\cite{Chivukula:2004af}
to compute the universal oblique and non-oblique corrections.
These are parameterized in terms of the leading parameters
$(\Shat,\,\That,\,W,\,Y)$ \cite{Barb}.
They are combinations of the parameters
$(S,\,T,\,\Delta\rho,\,\delta)$ \cite{Chivukula:2004af},
%
\begin{eqnarray}
\hat{S} \,=\,
\frac{1}{4s_W^2}
\left[\alpha S+4c_W^2(\Delta\rho-\alpha T)+\frac{\alpha\delta}{c_W^2}\right] ,
~~~~~~ \hat{T} \,=\, \Delta\rho \,,
~~~~~~
W \,=\, \frac{\alpha\delta}{\,4s_W^2c_W^2\,} \,,
~~~~~~
Y \,=\, \frac{c_W^2}{s_W^2}\(\Delta\rho-\alpha T\) \,.
\end{eqnarray}
%

For the gauge sector, with systematical calculations we derive,
\beq
\label{eq:ST-gauge}
\ba{lcl}
\alpha S_{\rm{g}} &\!\!=\!\!&
\dis\sww \left[ -4x^{4}y^{2} + 8x^{6}y^{2} - 4x^{6}y^{4}\(1+c_W^{-2}\)
+ \mathcal{O}(x^{8})\right] ,
\\[3mm]
\alpha T_{\rm{g}} &\!\!=\!\!&
\dis{\sww}{c_W^{-2}}
\left[ -2x^{6}y^{4} + \mathcal{O}(x^{8}) \right] ,
\\[3mm]
\alpha \delta_{\rm{g}}^{} &\!\!=\!\!&
\dis 4\sww\cww \left[x^4y^2 - 2x^6y^2 + 2x^6y^4
+ \mO(x^8) \right] ,
\ea
\eeq
as well as $\,\Delta\rho = 0\,$.\,
Thus, we arrive at
\beqa
\Shat_{\rm{g}} \,=\, \mO(x^6y^4)\,, ~~~~~~
\That_{\rm{g}} \,=\, 0 \,, ~~~~~~
W_{\rm{g}} \,=\, x^4y^2 + \mO(x^6y^2) \,, ~~~~~~~
Y_{\rm{g}} \,=\, \mO(x^6y^4) \,,
\eeqa
where we see that only $\,W_{\rm{g}}\,$ could be sizable, and
\,$(\Shat_{\rm{g}},\,Y_{\rm{g}})$\, are further suppressed by a factor of
$\,\mO(x^2y^2)\sim \mO(10^{-2})\,$.\,
From the Higgs and fermion sectors,
their leading non-oblique corrections are negligible at one-loop.
Hence, we just compute the leading oblique contributions to \,$(S,\,T)$.\,
In the Higgs sector, we have two neutral states \,$(h^0,\,H^0)$\, with
the mixing angle $\,\alpha$\,.\, Thus, we infer the oblique contributions,
\beq
\ba{lcl}
S_{\rm{s}} \!\!&=&\!\! \dis\frac{1}{12\pi}
      \left[ c^2_\alpha \ln\!\frac{M^2_h}{M_Z^2}\!
         - \ln\!\frac{(M^2_h)^{\text{sm}}_{\text{ref}}}{M_Z^2}\!
      + s^2_\alpha \ln\!\frac{M^2_H}{M^2_Z}\!\right]  \, ,
\\[5mm]
T_{\rm{s}} \!\!&=&\!\! \dis\frac{-3}{16\pi c_W^2}
     \left[ c^2_\alpha \ln\!\frac{M^2_h}{M_Z^2}\!
        - \ln\!\frac{(M^2_h)^{\text{sm}}_{\text{ref}}}{M_Z^2}\!
     + s^2_\alpha \ln\!\frac{M^2_H}{M^2_Z}\!\right]\,,
\ea
\eeq
where \,$(s_\al^{},\,c_\al )\equiv (\sin\al,\,\cos\al)$\,.
For the fermion sector with seesaw rotations (\ref{eq:ss-angle}),
systematical calculations give \cite{He:1999vp},
\beq
\ba{lcl}
S_{\rm{f}}^{} \!\!&=&\!\!
\dis\frac{4N_c}{9\pi} \left[\ln\!\frac{M_\mT^{}}{m_t^{}} - \frac{7}{8}
 + \frac{1}{16h_t^{}} \right] \frac{z_t^2}{\,1\!+ r\,} \,,
\\[5mm]
T_{\rm{f}}^{} \!\!&=&\!\!
\dis\frac{N_c h_t^{}}{16\pi s_W^2c_W^2}
\left[8\ln\!\frac{M_\mB^{}}{m_t^{}}\!
  +\!\frac{4}{3r} - 6 \right]
\frac{z_t^2}{\,1\!+ r\,} \,,
\ea
\eeq
where $\,h_t^{}\equiv m_t^2/M_Z^2\simeq 3.6\,$ and
$\,z_t^{} \equiv m_t^{}/\kappa \ll 1\,$.\,
We see that due to $\,z_t^{} = \mO(m_t^{}/M_S^{})\ll 1\,$,\, the fermionic contributions
can be fairly small and under control. This decoupling nature is because the heavy spectator
quarks $\,(\T,\,\B)\,$ are {\it vector-like} under $\,SU(2)_L^{}\,$,\, unlike the case
of a conventional fourth chiral family added to the SM \cite{Chen:2012wz}.
In passing, we note that Ref.\,\cite{Dawson} recently studied certain vector-fermion models
and their phenomenologies in a different context.

\begin{figure}[t]
\begin{center}
\includegraphics[width=8cm,height=6.3cm]{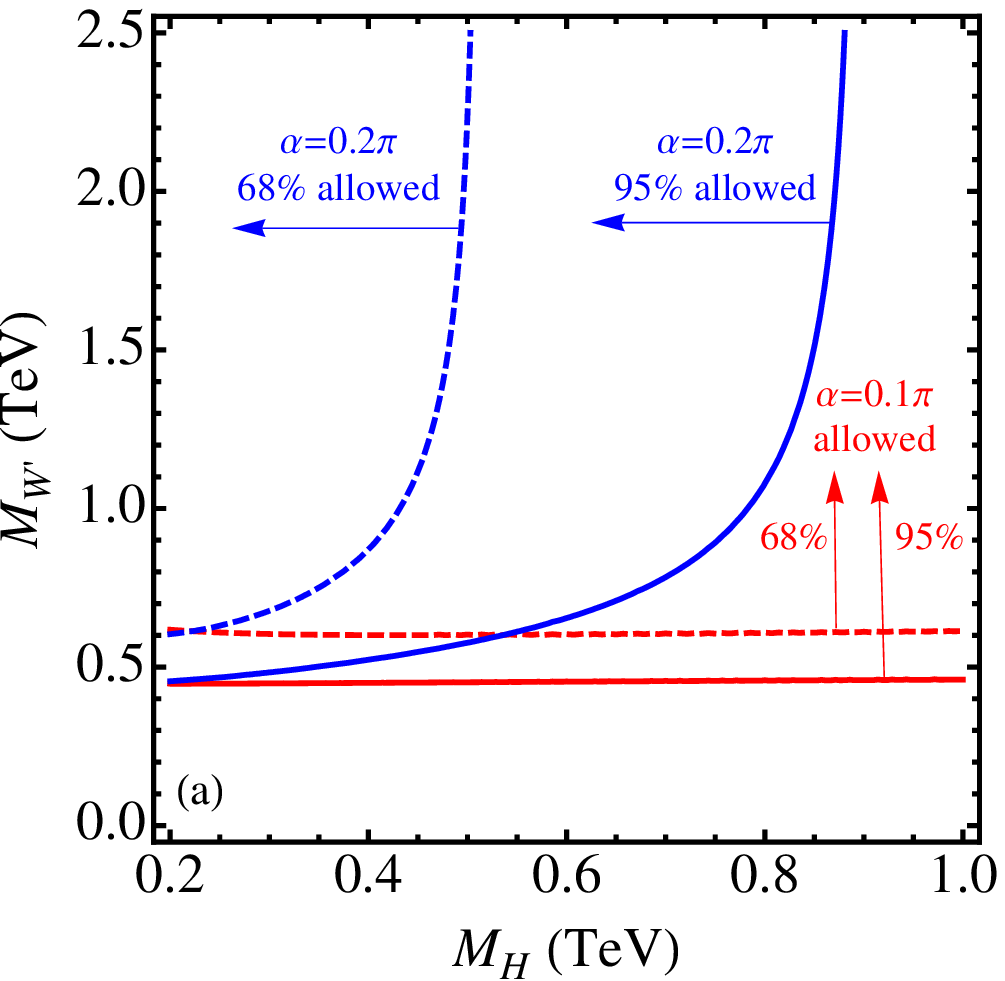}
\hspace*{2mm}
\includegraphics[width=8cm,height=6.3cm]{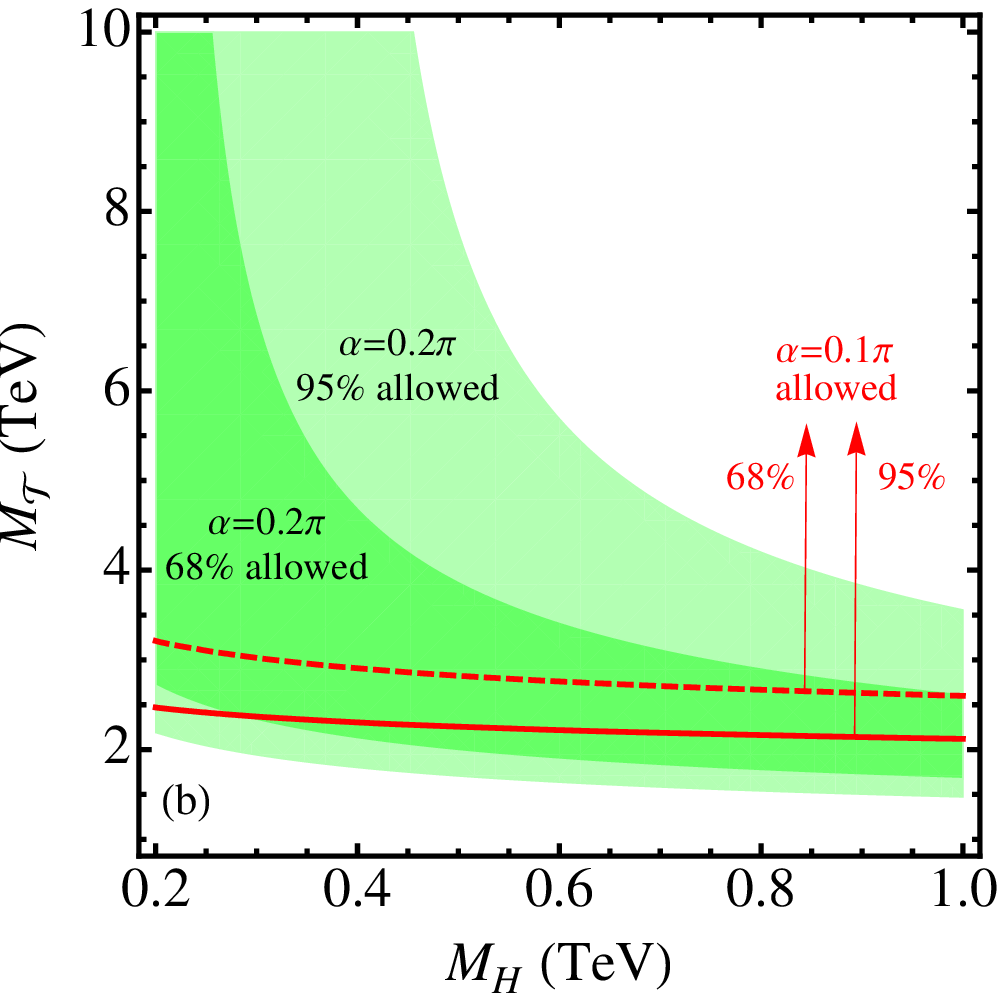}
\vspace*{-2mm}
\caption{
 Precision constraints on $\,W'$\, and $\,\T\,$ masses
 as functions of $\,H^0$\, mass.
 Plots (a) and (b) display the allowed ranges of parameter space
 in $\,M_H^{}-M_{W'}^{}$\, plane (with \,$M_{\T}^{}=4$\,TeV) and
 in the $\,M_H^{}-M_{\T}^{}$\, plane (with \,$M_{W'}^{}=1.5$\,TeV), respectively.
 In both plots, we have sample inputs $\,(x,\,r)=(0.2,\,1)$\,.
}
\label{fig:STfit}
\end{center}
\end{figure}

Summing up the above contributions from gauge, Higgs and fermion
sectors, we deduce the predicted total $(\Shat,\,\That,\,W,\,Y)$ and
compare them with the electroweak precision fit \cite{Barb}.  With these, we
can derive the constraints on our model. In Fig.\,\ref{fig:STfit}, we
present the 68\% and 95\% confidence limits on the allowed ranges of our parameter space.
Plot-(a) displays the allowed space for $\,M_H^{}$\, versus \,$M_{W'}^{}$,\,
with the sample inputs $\,\al = 0.1\pi,\,0.2\pi\,$ and $\,M_{\T}^{}=4$\,TeV.\,
We see that the 95\% confidence limits only require
$\,M_{W'}^{} \gtrsim 0.45-1\,$TeV for wide $H^0$ mass range up to 800\,GeV and
mixing angle $\,\al = (0.1-0.2)\pi\,$.\,
Plot-(b) depicts the viable parameter region in the $\,M_H^{}-M_{\T}^{}$\,
plane, where we input $\,\al = 0.1\pi,\,0.2\pi\,$ and \,$M_{W'}=1.5$\,TeV\,.\,
It shows that the heavy quarks $\,\T\,$ (and $\,\B\,$)
should have mass above $\,1.5-2.5$\,TeV at 95\%\,C.L.
In both plots, we have sample inputs $\,(x,\,r)=(0.2,\,1)$, consistent
with the global fit in Sec.\,3.\,

We note that the low energy Fermi constant $\,G_F^{}$\, in our model
is derived from the charged current with exchanges of
$\,W\,$ and $\,W'\,$ bosons in the zero momentum limit,
\beqa
4\sqrt{2}G_{F} ~=~
\frac{\,G^{2}_{Wff}}{M^{2}_{W}\,}+\frac{\,G^{2}_{W'ff}\,}{M^{2}_{W'}}\,,
\eeqa
where $\,G^{}_{Wff}\,$ and $\,G^{}_{W'ff}\,$
stand for the gauge couplings of $\,W\,$ and $\,W'\,$ with
the light fermions [except $\,(t,\,b)$\, and heavy spectator quarks],
respectively. Analyzing the diagonalization of the mass-matrix
for charged gauge bosons in Eq.\,(\ref{eq:WZ-MassM}), we can
generally prove,
\beqa
\label{eq:GF-v}
\sqrt{2}G_{F}^{} ~=~
\frac{1}{4}g_1^{2}\(\mathbb{M}^{-2}_{W}\)_{22}^{}
\,=~ \frac{1}{\,v^2\,}\,.
\eeqa
This shows that $G_F^{}$ receives no extra correction in the present model.
Similarly, we find that no new correction to the neutral current process in the
zero-momentum limit, and thus  $\,\Delta\rho=0$\,.\,

We also analyze the electroweak measurement via
neutrino-nucleon scattering $\,\nu_{\mu}^{} N\to\nu_{\mu}^{} N$\,.\,
It is one of the most precise probes of the weak neutral current and is
not included in the global fit of $\,(\Shat,\,\That,\,W,\,Y)$ \cite{Barb}.\,
The effective Lagrangian for weak neutral current of $\,\nu-q\,$ scattering
is given by
\beqa
\mathcal{L} ~=~
-\frac{G_{F}}{\sqrt{2}}\[\bar{\nu}\gamma^{\mu}(1-\gamma^{5})\nu\]
\[\epsilon^{q}_{L}\bar{q}\gamma_{\mu}^{}(1-\gamma^{5})q
  +\epsilon^{q}_{R}\bar{q}\gamma_{\mu}^{}(1+\gamma^{5})q \] ,
\eeqa
where the isoscalar combination
$\,g^{2}_{L,R}=(\epsilon^{u}_{L,R})^{2}+(\epsilon^{d}_{L,R})^{2}$\,
is measured by the NuTeV collaboration \cite{NuTeV},
$\,(g^{\textrm{eff}}_{L})^{2}=0.3005\pm 0.0014$\,,\,
about $\,2.6\sigma\,$ lower than the SM prediction
$\,(g_L^2)_{\text{sm}}^{}=0.3042$\,.\,
In our model, we derive the four-fermion operator for $\,\nu - q\,$ scattering process
with zero-momentum transfer,
\beqa
\left.\mathcal{M}[\nu q\to\nu q]\right|_{q^{2}\to 0}^{}
~=~-\frac{\,G_{Z\nu\nu}^{}G_{Zqq}^{}\,}{M^{2}_{Z}}
   -\frac{\,G_{Z'\nu\nu}^{}G_{Z'qq}^{}\,}{M^{2}_{Z^{'}}} \,,
\eeqa
where $\,G_{Z\nu\nu}^{}\,$ ($\,G_{Zqq}^{}\,$) and $\,G_{Z'\nu\nu}^{}\,$ ($\,G_{Z'qq}^{}\,$)
represent the gauge couplings of neutrinos (light quarks) with $Z$ and $Z'$, respectively.
Extracting the $\,\epsilon^{u,d}_{L,R}\,$ parameters, we compute the effective coupling,
$\,g^{2}_{L,R}= (g_{L,R}^2)_{\text{sm}}^{}[1+\mO(x^8)]\,$.\,
Hence, our prediction agrees well with the SM. This is unlike the early non-universality
model (NUM)\,\cite{NU} which assigns a different $SU(2)$ for the first two families of
light fermions and leads to,
$\,g^2_L/(g^{}_L)^2_{\text{sm}} \simeq 1+2.57(v/u)^2$\,.\,
This sizable correction severely constrains the VEV $\,u$\, and pushes
\,$W'/Z'$\, masses above 3.6\,TeV \cite{x2010} for the NUM.

Next, we analyze the direct search limits on the new gauge bosons
$(W',\,Z')$. For this purpose, we first derive
the trilinear couplings of $W'$ and $Z'$ with the light fermions,
top/bottom quarks, light gauge bosons and Higgs bosons.
For convenience, we define the ratios of $\,V'$ ($=W',Z'$)\, couplings over that of the
light $\,V$ ($=W,Z$)\, boson in the SM,
\beqa
\xi_{V'ff}^{}\,=\,\frac{G_{V'ff}^{}}{G_{Vff}^{\text{sm}}}\,,~~~~~~
\xi_{V'VV}^{}\,=\,\frac{G_{V'VV}^{}}{G_{VVV}^{\text{sm}}}\,,~~~~~~
\xi_{V'Vh}^{}\,=\,\frac{G_{V'Vh}^{}}{G_{VVh}^{\text{sm}}}\,,~~~~~~
\xi_{V'VH}^{}\,=\,\frac{G_{V'VH}^{}}{G_{VVh}^{\text{sm}}}\,,
\eeqa
where the subscript $f$ stands for SM fermions.
We expand these coupling ratios in terms of $\,(x,\,y)\,$
and summarize them in Table\,\ref{tab2:coup-V'XY}.

\begin{table}[t]
\begin{center}
\renewcommand{\arraystretch}{1.3}
\begin{tabular}{c|c|c|c|c|c|c}
\hline\hline
$\xi_{W'ff}^{}$ & $\xi_{W'tb}^{}$ & $\xi_{W'WZ}^{}$ &
$\xi_{W'Wh}^{}$ & $\xi_{W'WH}^{}$ & $\xi_{Z'u_{L}u_{L}}^{}$ & $\xi_{Z'd_{L}d_{L}}^{}$
\\[1mm]
\hline
$-x$ & $\frac{1-rx^{2}}{x(1+r)}$ & $\frac{x^{3}y^{2}}{c^{2}_{W}}$ & $-x(\ca\!-\!y\sa)$ & $x(\sa\!+\!y\ca)$ & $-x\cw/(1-\frac{4}{3}\sww)$ & $-x\cw/(1-\frac{2}{3}\sww)$
\\[1mm]
\hline\hline
$\xi_{Z't_{L}t_{L}}^{}$ & $\xi_{Z't_{R}t_{R}}^{}$ &
$\xi_{Z'b_{L}b_{L}}^{}$ & $\xi_{Z'b_{R}b_{R}}^{}$ &
$\xi_{Z'WW}^{}$ & $\xi_{Z'Zh}^{}$ & $\xi_{Z'ZH}^{}$
\\[1mm]
\hline
$\frac{\cw}{x(1+r)(1-\frac{4}{3}\sww)}$ & $-\frac{3\cw z^{2}_{t}}{4x\sww(1+r)}$ & $\frac{\cw}{x(1+r)(1-\frac{2}{3}\sww)}$ &
$-\frac{3\cw z^{2}_{b}}{2x\sww(1+r)}$ & $\frac{x^{3}y^{2}}{\cw}$ & $-x\cw(\ca\!-\!y\sa)$ & $x\cw(\sa\!+\!y\ca)$
\\[1mm]
\hline\hline
\end{tabular}
\end{center}
\vspace*{-3mm} %
\caption{Gauge coupling ratios $\,\xi_{W'XY}^{}\,$ and $\,\xi_{Z'XY}^{}\,$
of $\,W'$\, and $\,Z'$\, bosons with the SM particles
$\,XY= f_1^{}\bar{f}_2^{},\,t\bar{b},\,t\bar{t},\,\bbbar,\,VV,\,Vh,\,VH\,$,\,
where $\,V=W,Z\,$,\, and $\,f_1^{}\bar{f}_2^{}$\, denote the light SM fermions other than
$\,(t,b)$\,.
}
\label{tab2:coup-V'XY}
\end{table}

The ATLAS and CMS collaborations have been actively searching for new gauge bosons
$W'$ and $Z'$ at the LHC \cite{V'-ATLAS}\cite{V'-CMS}.  They mainly focus on the
sequential standard model (SSM), where the couplings of \,$W'$\, and \,$Z'$\,
with fermions equal the corresponding SM couplings of light $W$ and $Z$ bosons.
But, our model essentially differs from the SSM.
As shown in Table\,\ref{tab2:coup-V'XY}, the predicted couplings of $\,W'/Z'$\, with
light fermions are suppressed by the small mixing angles between \,$V_0^{}$\, and \,$V_1^{}$,\,
which are of $\,\mO(x)$\,.\, Hence, the production rates of $\,W'$\, and $\,Z'$\, are proportional
to $\,x^{2}\,$,\, and thus much harder to detect. On the other hand, the couplings
of $\,W'/Z'$\, with top and bottom quarks are enhanced by the factor $\,\frac{1}{x}\,$
(cf.\ Table\,\ref{tab2:coup-V'XY}). This means that the decay branching fractions of
$\,W'\to tb\,$ and $\,Z'\to t\bar{t},\,b\bar{b}\,$ are enhanced.
ATLAS already searched for leptonic decay modes
$\,W'\to \ell\nu\,$ and $\,Z'\to \ell^+\ell^-\,$ (with $\ell = e,\mu$) \cite{V'-ATLAS},
while CMS explored the quark decay channels
$\,W'\to t\bar{b},\,\bar{t}b\,$ and $\,Z'\to t\bar{t}\,$ \cite{V'-CMS}.

To compare the LHC experimental search limits (based on SSM hypothesis) with
our theory predictions, we will rescale the SSM cross sections and branching fractions
to our model. For $\,pp\to W'\to tb\,$ channel,
CMS explicitly gives $W'_{R}$ search limit \cite{V'-CMS} at the present.
We note that our case of $\,W'=W'_L$\, is rather similar.
Although the signal process $\,pp\to W'_L\to tb\,$
has interference with the SM process $\,pp\to W\to tb\,$,\,
for heavy $W'_L$ with mass above $600-800$\,GeV, the interference
is fairly small around the $W'_L$ mass window.
Hence, for the estimate we may directly rescale the $W'_{R}$ search limits \cite{V'-CMS}
for our constraints. The SSM hypothesis takes $W'$ couplings with all SM fermions
equal the corresponding SM couplings of $W$, and for heavy $W'$ one can easily find,
\,Br$[W'\to tb]\simeq \frac{1}{4}$\, \cite{V'-CMS}.
In our model, Table\,\ref{tab2:coup-V'XY} shows that $W'$ couplings with light fermions
are suppressed by $\,\xi_{W'ff}^{}=-x\,$,\, and its coupling with $\,tb$\, is enhanced
by a sizable factor of $\,\frac{1}{x}\,$.\, Thus, we readily deduce
\,Br$[W'\to tb]\simeq 1$\, in the present model.
With these we find that our $W'$ signal rate
of $\,\sigma\times\text{Br}\,$ is smaller than that of the SSM \cite{V'-CMS} by about
a factor of $\,4x^2\,$.\,

For numerical estimate, we find the 95\%\,C.L.\ lower limit
on $W'$ mass, $\,M_{W'}^{} > 1.25$\,TeV, from the CMS data \cite{V'-CMS} and
with the sample input $\,x=0.2\,$.\,
This mass limit becomes stronger when the parameter $\,x=g_{1}/g_{0}\,$ increases.
For instance, inputting $\,x=0.25\,$,\, we have,
$\,M_{W'}^{} > 1.6$\,TeV.
Similarly, we analyze the process $\,pp\to W'\to \ell\nu$\,
measured by ATLAS \cite{V'-ATLAS}
and obtain a weaker limit. The $Z'$ gauge boson can be probed via
$\,pp\to Z'\to \ell\ell,\, t\bar{t}\,$.\, We find that CMS gives a stronger limit,
$\,M_{Z'}^{} > 1.0\,$TeV at 95\%\,C.L.,
via the process $\,pp\to Z'\to \ell\bar{\ell}\,$ ($\ell = e,\mu$).\,
In passing, the LHC detections of certain $W'/Z'$ bosons were recently
considered for different models\,\cite{Du:2012vh}.


\vspace*{2mm}
\section{Lighter Higgs Boson Signals at the LHC}
\vspace*{2mm}

In this section, we analyze the production and decays of
the 125\,GeV Higgs boson $\,h^0$\, at the LHC.
With these, we compare our predictions with the Higgs searches at the LHC.
We first perform a fit with the ATLAS and CMS data, and then make a global fit
by further including the Tevatron data and precision constraints.
From this, we determine the favored parameter space of our theory.

To analyze the gauge and Yukawa couplings of $\,h^0\,$ and $\,H^0\,$,\,
it is convenient to define the ratios,
\beqa
\xi_{h\alpha\beta}^{}
~=~ \frac{\,G_{h\alpha\beta}^{}\,}{\,G_{h\alpha\beta}^{\text{sm}}\,}\,,
~~~~~~
\xi_{H\alpha\beta}^{}
~=~ \frac{\,G_{H\alpha\beta}^{}\,}{\,G_{h\alpha\beta}^{\text{sm}}\,}\,,
~~~~~~
\xi_{hX\bar{X}}^{} ~=~ \frac{\,v\,}{\,M_X^{}\,}G_{hX\bar{X}}^{}\,,
~~~~~~
\xi_{HX\bar{X}}^{} ~=~ \frac{\,v\,}{\,M_X^{}\,}G_{HX\bar{X}}^{}\,,
\eeqa
where $\,\alpha\beta =f\bar{f},\,b\bar{b},\,t\bar{t},\,VV\,$
and $\,X\bar{X}=\T\bar{\T},\,\B\bar{\B}\,$.\,
With systematical analyses, we present Higgs couplings with gauge bosons and fermions
in Table\,\ref{tab3:Hig-coup}.  We see that  $\,h^0$\, couplings with the
SM fermions and gauge bosons are smaller than the corresponding SM values
due to the Higgs mixing factor $\,\ca < 1\,$,\, and $\,h^0$\, couplings
with heavy fermions and new gauge bosons are generally suppressed by a coefficient
$\,y\sa\,$,\, where $\,(\sa ,\,\ca )=(\sin\alpha,\,\cos\alpha)<1\,$
and $\,y=v/u \ll 1\,$.\,  The $h^0$ couplings with $\,t\bar{t}\,$ and $\,\bbbar\,$
receive an additional shift of $\,\mO(y\sa)\,$.

\begin{figure}[t]
\vspace*{3mm}
\begin{center}
\includegraphics[width=7.8cm,height=5.85cm]{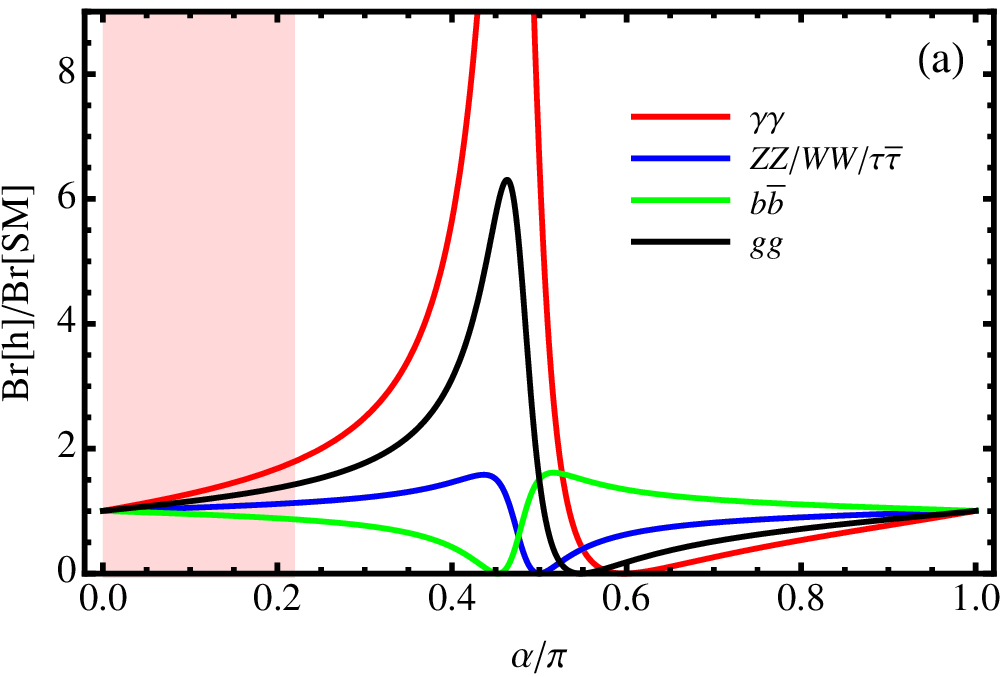}
\hspace*{2mm}
\includegraphics[width=8cm,height=6cm]{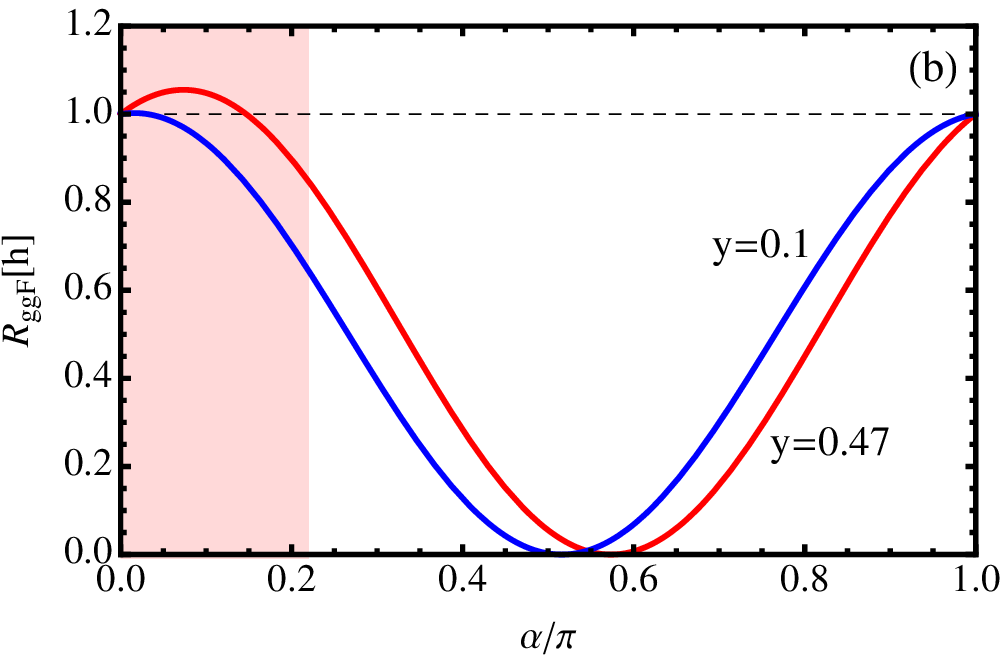}
\vspace*{-3.5mm}
\caption{Decays and productions of the Higgs boson $\,h^0\,(125\,\GeV)\,$ at the LHC.
Plot-(a) shows the ratios of $h^0$ decay branching fractions over that of
the corresponding SM values, as a function of Higgs mixing angle $\,\alpha\,$
and for the major decays channels. 
Plot-(b) depicts the ratio of $\,gg\to h^0$\, production cross section over the SM value.
In plot-(a) we set $\,(r,\,y)=(1,\,0.29)$\, and
$\,(M_{W'}^{},\,M_{\T}^{})=(1.4,\,4)$TeV.\,
In plot-(b) we have sample inputs
$\,(r,\,M_{\T}^{})=(1,\,4\,\TeV)\,$,\, as well as
$\,y=0.47~(0.1)$\, for red (blue) curve
from our best fit with ATLAS (CMS) data in Fig.\,\ref{fig5:y-a-bound}(a).
The shaded pink region in each plot is the $1\sigma$ favored range
of the mixing angle $\,\alpha$\,,\, as given by our global fit
in Fig.\,\ref{fig5:y-a-bound}(b).
}
\label{fig:2}
\end{center}
\end{figure}

Then, we analyze decays and productions of the Higgs boson $h^0$ with mass 125\,GeV.
Since $h^0$ couplings with SM particles are mainly suppressed by $\,\ca$\,,\,
its total decay width roughly decreases with $\,c_\al^2\,$.\,
Fig.\,\ref{fig:2}(a) shows the ratios of the $h^0$ decay branching fractions over the
corresponding SM values, for the major decay channels.
We see that \,Br$[h\to\gamma\gamma]$\, decreases rapidly around
$\,\al = \frac{\pi}{2}\,$ due to the factor $\,c_\al^2\,$,\,
but gets enhanced over the SM values in the region
$\,0< \alpha < \frac{\pi}{2}$\,.\,
This enhancement is interesting because all the Higgs couplings are suppressed for
$\,\alpha \in \(0,\, \frac{\pi}{2}\)$\, range, and thus the Higgs partial widths
and total width become smaller than the SM values.
So, naively we do not expect an enhancement here.
But,  smaller partial widths do not necessarily imply smaller decay branding fractions,
because the total decay width also reduces accordingly.
We note that the total decay width is mainly contributed by
$\,h^0\to b\bar{b}\,$ channel, which however receives more suppression than other
channels. As shown in Table\,\ref{tab3:Hig-coup}, the $\,h^0b\bar{b}\,$
coupling receives an extra negative reduction of $\,\mO(-y\sa)\,$, which makes
the total width reduce more than the partial widths of $\,\gamma\gamma\,$,
$\,gg\,$ and $\,WW/ZZ/\tau\tau\,$ over the range of
$\,\alpha \in \(0,\, \frac{\pi}{2}\)$\,.\,
Besides, the diphoton partial width is dominated by the $W$-loop (with a
$\,c_\al^2\,$ suppression), and adding the $W'$-loop partly delays this suppression
since $W'$-loop induces a term of $\,\mO(y^2s_\al^2)$\,.\,
We further note that \,Br$[h\to gg]$\, has extra contributions from the
heavy quark $\,\T/\B$\, triangle-loops, where the $h^0$ Yukawa couplings
with $\,\T\,(\B)$\, are dominated by the factor $\,y\sa\,$
(Table\,\ref{tab3:Hig-coup}).
In consequence, we find that the decay branching fractions of $\,\gamma\gamma\,$
and $\,gg\,$ are significantly enhanced in the region of
$\,\alpha \in \(0,\, \frac{\pi}{2}\)$\,,\,
as demonstrated in Fig.\,\ref{fig:2}(a).

\begin{table}
\vspace*{3mm}
\begin{center}
\renewcommand{\arraystretch}{1.3}
\begin{tabular}{c||c|c|c|c|c|c|c}
\hline\hline
&&&&&&& \\[-4mm]
$XY$	 & $f\bar{f}$ & $VV$ & $V'V'$ & $t\bar{t}$ & $b\bar{b}$
& $\T\bar{\T}$ & $\B\bar{\B}$
\\
\hline\hline
&&&&&&& \\[-4mm]
$\xi_{hXY}^{}$ & $\ca$ & $\ca$ & $y\sa$ & $\ca\!-\!\dis\frac{y\sa}{1\!+\!r}$ & $\dis\ca\!-\!\frac{y\sa}{1\!+\!r}$ &
$\dis\frac{\,y\sa\!+\omega_h^{}z^{2}_{t}\,}{1\!+\!r}$ &
$\dis\frac{y\sa}{1+r}$
\\[2mm]
\hline
&&&&&&& \\[-4mm]
$\xi_{HXY}^{}$ & $-\sa$ & $-\sa$ & $y\ca$ &
$-\sa\!-\!\dis\frac{y\ca}{1\!+\!r}$ & $-\dis\sa\!-\!\frac{y\ca}{1\!+\!r}$ & $\dis\frac{\,y\ca\!-\omega_H^{}z^{2}_{t}\,}{1\!+\!r}$ &
$\dis\frac{y\ca}{1+r}$
\\[2mm]
\hline\hline
\end{tabular}
\end{center}
\vspace*{-3mm}
\caption{Coupling ratios $\xi_{hXY}^{}$ and $\xi_{HXY}^{}$
of Higgs boson $h^0$ and $H^0$ with the SM particles
$\,XY= f\bar{f},\,VV,\,V'V',\,t\bar{t},\,b\bar{b},\,\T\bar{\T},\,\B\bar{\B}\,$,\,
where $\,V=W,Z\,$ and $\,V'=W',Z'\,$.\, We also denote,
$\,\omega_{h}^{} \equiv \ca - y\sa (1-r)/(1+r)$\, and
$\,\omega_{H}^{} \equiv \sa + y\ca (1-r)/(1+r)$\,.
}
\label{tab3:Hig-coup}
\end{table}

The major channel of $h^0$ productions at the LHC comes from the gluon fusions.
For the LHC\,(7+8\,TeV) and LHC\,(14\,TeV),
about 87\% of the Higgs boson events are produced in this process.
The ratio of our $h^0$ production cross section over that of the SM Higgs boson
with the same mass is derived as,
\beqa
\mR_{ggF}^{}[h] ~=~ \frac{\sigma [{gg\to h}]}{\,\sigma[{gg\to h}]_{\rm{SM}}^{}\,}
~=~ {\,\left|\sum\limits_{q=t,\mT,\mB}
          \xi_{hqq}^{}A^{H}_{1/2}(\tau_{q})\right|^{2}}\,
    {\left|\sum\limits_{q=t}A^{H}_{1/2}(\tau_{q})\right|^{-2}}\,,
\label{eq:RggF-h}
\eeqa
where $\,\tau_{q}\equiv M^{2}_{h}/(4m^{2}_{q})$\,,\,
and the fermion-loop form factor is given by
\begin{subequations}
\begin{eqnarray}
&&
A_{1/2}^H(\tau) ~=~ 2[\tau+(\tau-1)f(\tau)]\tau^{-2}\,,
\\[2mm]
&&
f(\tau) ~=~ \begin{cases}
~ \arcsin^2\!\!\sqrt{\tau} \,\,, & \tau\leqq 1\,,
\\[2.5mm]
~ -\frac{1}{4}\!
\left[\displaystyle\ln\frac{1+\sqrt{1-\tau^{-1}}}{1-\sqrt{1-\tau^{-1}}}-i\pi\right]^2\!,~~
 & \tau>1 \,.
\end{cases}
\end{eqnarray}
\end{subequations}
Under the heavy fermion mass limit $\,M_h^2\ll 4m_q^2$\,,\,
the function $\,A_{1/2}^H\,$ takes asymptotic form,
$\,A_{1/2}^H(\tau)\to \frac{4}{3}$\,.\,
This means that loop form factors
of the new spectator quarks are largely the same as the top quark.

Thus, as an estimate we can approximate the production rate as follows,
\beqa
\mR_{ggF}[h]
~\simeq~ \(\xi_{htt}^{}+\xi_{h\mT\mT}^{}+\xi_{h\mB\mB}^{}\)^{2}
\,\simeq~ \(\ca+y\frac{\sa}{\,1\!+\!r\,}+z^{2}_{t}\frac{\omega_h^{}}{\,1\!+\!r\,}\)^{2}
\,\simeq~ \(\ca + \frac{y\sa}{\,1\!+\!r\,}\)^2  \,,
\label{eq:RggF-app}
\eeqa
where $\,z_t^2 \ll y \ll 1 \,$.
In Fig.\,\ref{fig:2}(b), we compute the production cross section ratio $\mR_{ggF}^{}$
as a function of the Higgs mixing angle $\,\alpha\,$.\,
It shows that this ratio is mainly suppressed by $\,\ca\,$,\, and the small
enhancement around $\,\alpha =0.1\pi\,$ arises from the $\,y\sa /(1\!+\!r)\,$ term
in (\ref{eq:RggF-app}). Hence, for small $\,y\,$ and $\,\al\,$,\,
we see that top quark loop gives the main contribution
to the production rate $\,\mR_{ggF}^{}$\,.\,

\begin{figure}[t]
\begin{center}
\includegraphics[width=8cm,height=6.5cm]{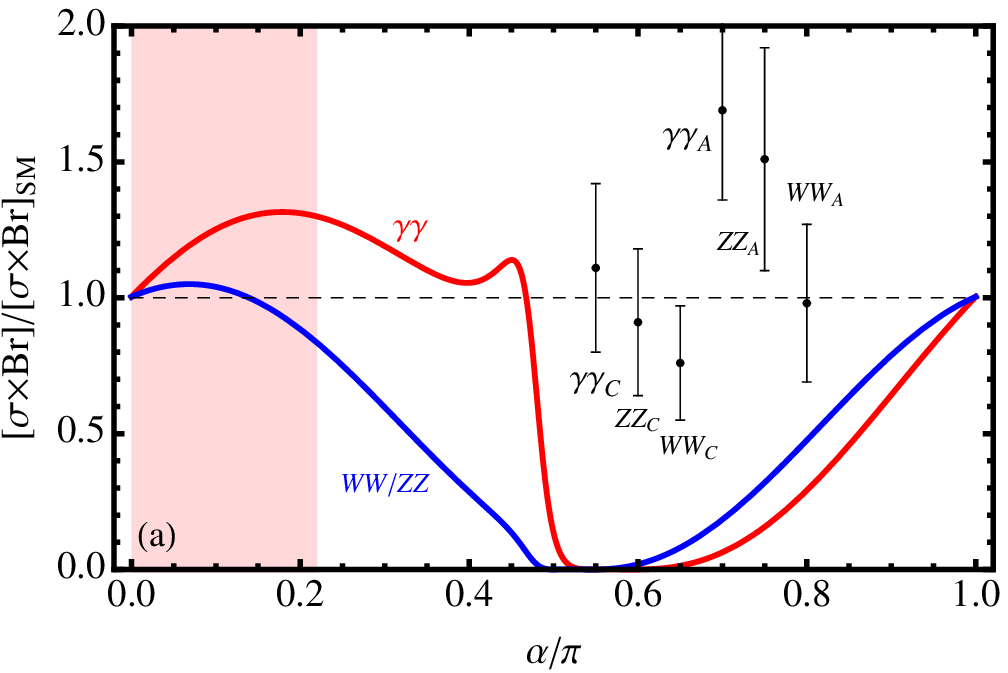}
\hspace*{2mm}
\includegraphics[width=8cm,height=6.5cm]{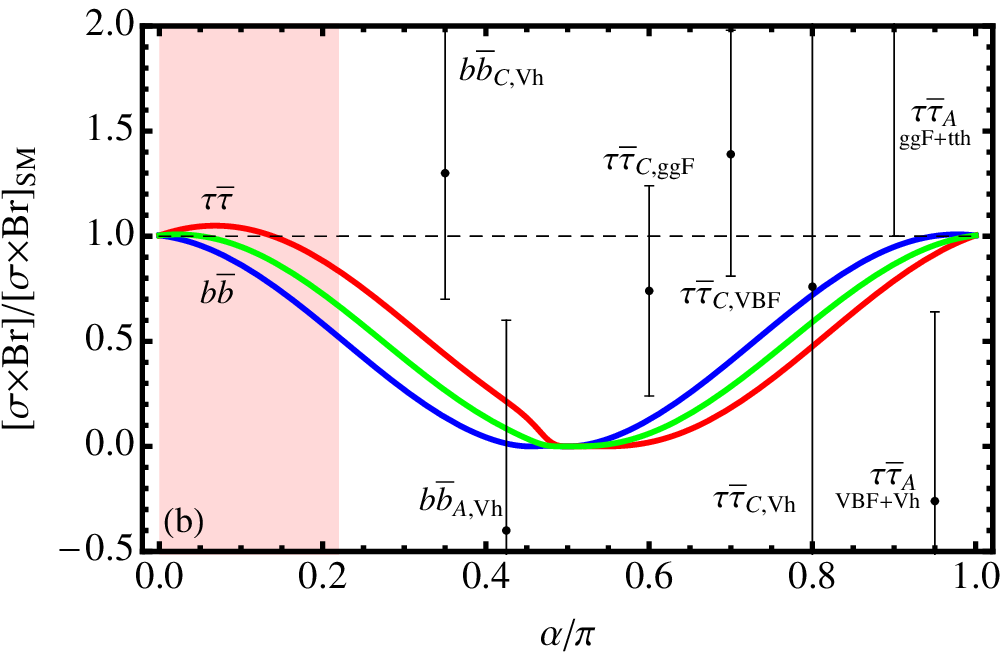}
\vspace*{-3mm}
\caption{Predicted LHC signal ratios
$\,[\sigma\times\text{Br}]/[\sigma\times\text{Br}]_{\text{SM}}^{}\,$
as functions of Higgs mixing angle $\,\alpha$\,.\,
Plot-(a) depicts signal rates of $\,h\to\gaga,\,WW^*,\,ZZ^*\,$ from gluon fusions.
Plot-(b) presents $\,h\to\tautaub,\,\bbbar\,$ signals via
gluon fusion ($gg$F), vector boson fusion (VBF),
and associate production ($Vh$) processes.
The red (blue) curve of plot-(a) corresponds to $\,\gamma\gamma$\, ($WW^*/ZZ^*$)
channels. In plot-(b), the red (green) curve shows $\,h^0\to \tau\bar{\tau}$\,
via gluon fusions (vector boson fusions),  while the blue curve depicts
$\,h^0\to b\bar{b}\,$ via associate productions.
We have sample inputs
$\,(y,\,r,\,M_{W'}^{},\,M_{\T}^{})=(0.29,\,1,\,1.4\,\TeV,\,4\,\TeV)$\,
based our best fit.
The latest ATLAS/CMS data are also shown,
where in each label the subscript ``$_A$'' (``$_C$'') denotes ATLAS (CMS).
These data points are independent of $\,\al$\,,\, and their horizontal locations
are arbitrarily chosen, for the convenience of presentation.
The shaded pink region depicts the $1\sigma$ favored range
of mixing angle $\,\alpha$\,,\, as given by our global fit
in Fig.\,\ref{fig5:y-a-bound}(b).
}
\label{fig3:R-VV}
\end{center}
\vspace*{-5mm}
\end{figure}

To contrast the LHC data, we will analyze the signal ratio of our prediction
over the SM expectation, for each given channel $\,gg\to h\to XX\,$,
\beqa
\mR_{XX}^{}[h] ~\equiv~ \dis
\frac{\sigma[gg\to h]\times\rm{Br}[h\to XX]}
     {\,\sigma[gg\to h]_{\rm{SM}^{}}\times\rm{Br}[h\to XX]_{\rm{SM}}^{}\,} \,.
\eeqa
In Fig.\ref{fig3:R-VV}(a), we present the signal ratios
$\,(\mR_{\gamma\gamma}^{},\,\mR_{WW}^{},\,\mR_{ZZ}^{})$\,
as functions of the Higgs mixing angle $\,\alpha\,$,\, for $h^0$ Higgs boson with
mass 125\,GeV. It shows that our model can predict enhanced diphoton rate over
significant parameter space of $\,\al\,$,\,  for the sample inputs
$\,(y,\,r,\,M_{W'}^{},\,M_{\T}^{})=(0.29,\,1,\,1.4\,\TeV,\,4\,\TeV)$\,.\,
At the same time, the predicted signals in $\,WW^*$\, and $\,ZZ^*$\, channels can be
quite close to the SM values, especially for $\,0 < \al \lesssim 0.2\pi\,$.\,
These agree well to the latest LHC data \cite{LHC2013-3}\cite{LHC2013-3b}.

Then, we study the vector boson fusion process
$\,pp\to h^0 jj\,$ with $\,h^0\to \tau\bar{\tau}\,$,\,
and  the associate production $\,pp\to h^0 V\,$ with
$\,h^0\to b\bar{b}\,$.\,
Given the Higgs couplings of Table\,\ref{tab3:Hig-coup},
we compute the predicted signal ratios over that of the SM
as the following,
\beqs
\label{eq:AP-VBF}
\beqa
\label{eq:AP}
\hspace*{-10mm}
\mR_{Vh}^{}[h] &\!\!\equiv\!\!&
\frac{\sigma[q_1^{}\bar{q}_2^{} \to\! V h] \times {\rm Br}[h \to\! f\bar{f}]}
  {\sigma[q_1^{}\bar{q}_2^{} \to\! Vh]_{\rm SM}\times {\rm Br}[h\to\! f\bar{f}]_{\rm SM}^{}}
~\simeq~  \xi_{hVV}^2 \xi_{hff}^2 \frac{\,\Gamma_h^{\rm SM}\,}{\,\Gamma_h^{}\,} \,,
\\[2mm]
\hspace*{-10mm}
\mR_{\text{VBF}}^{}[h] &\!\!\equiv\!\!&
\frac{\sigma[q_1^{} q_2^{} \to h q_3^{}q_4^{}] \times {\rm Br}[h\to\! f\bar{f}]}
  {\sigma[q_1^{} q_2^{} \to h q_3^{}q_4^{}]_{\rm SM}^{}
   \times{\rm Br}[h \to\! f\bar{f}]_{\rm SM}^{}}
~\simeq~  \xi_{hVV}^2 \xi_{hff}^2 \frac{\,\Gamma_h^{\rm SM}\,}{\,\Gamma_h^{}\,}
\,,~~
\label{eq:VBF}
\eeqa
\eeqs
where the quarks $\,q_{1,2}^{}$\, or $\,q_{3,4}^{}$\, are all light quarks.
The involved $\,Vq\bar{q}'$\, couplings agree with the SM values
to good precision, and their deviations from the SM value arise
only at $\,\mO(x^{4}y^{2})$\, which is negligible.
Hence, the ratios of production cross sections for both the vector boson fusion and
associate production equal $\,\xi_{hVV}^2$\,.\,

\begin{figure}
\begin{center}
\includegraphics[width=7.8cm,height=6.8cm]{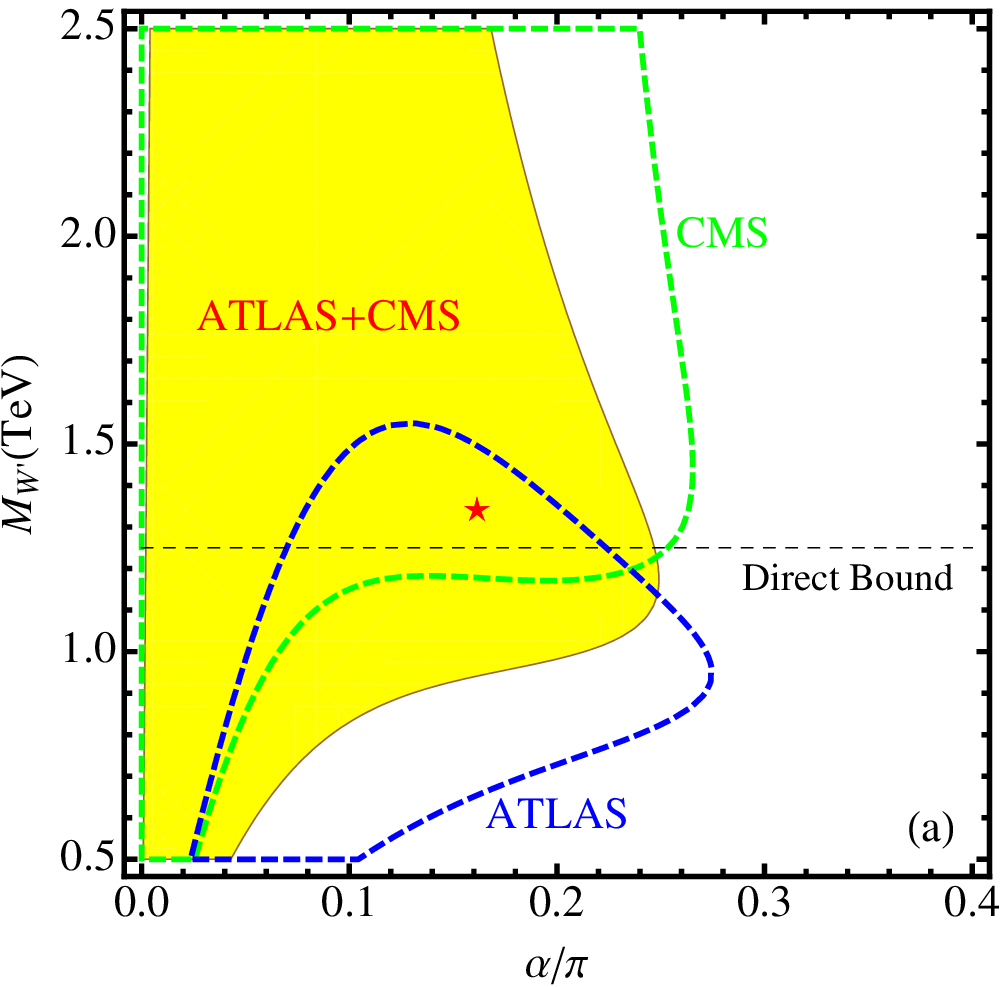}
\hspace*{2mm}
\includegraphics[width=7.8cm,height=6.8cm]{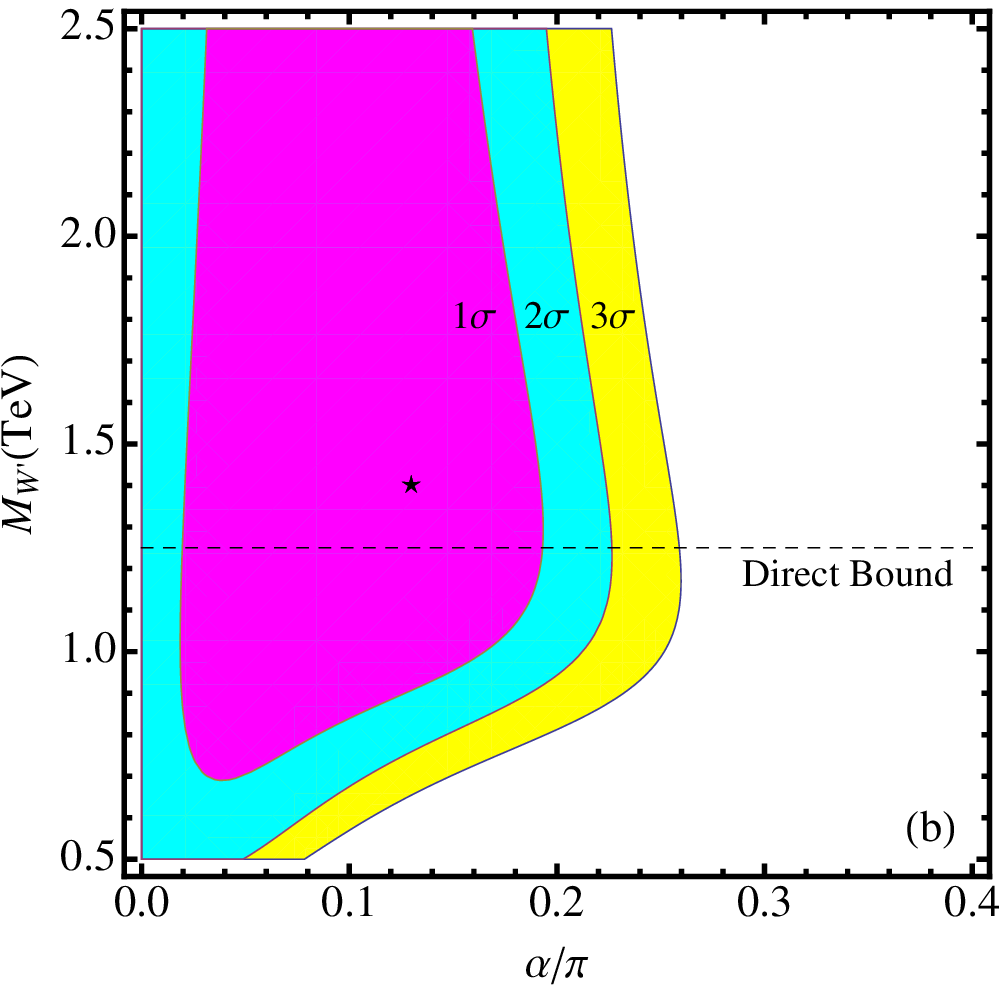}
\vspace*{-3mm}
\caption{Global fits for constraints on the allowed regions in
$\,\al - M_{W'}^{}\,$ plane.
Plot-(a) depicts the 68\%\,C.L.\ bound by fitting all the current ATLAS and CMS data,
where the yellow contour is ATLAS+CMS combined limit.
Plot-(b) presents the global fit by including all direct searches
of \,(ATLAS, CMS, Tevatron)\, and the indirect precision data,
at $\,(1\sigma,\,2\sigma,\,3\sigma)\,$ levels, as marked by
(pink, blue, yellow) contours, respectively.
In each plot, the red (black) star indicates the best fit point, and
the horizontal dashed line gives the 95\%\,C.L.\ lower bound,
$M_{W'}^{}> 1.25$\,TeV, from the LHC direct searches of $W'$ (Sec.\,2.2).
We have sample input \,$(x,\,r)=(0.2,\,1)$\, and
$\,(M_H^{},\,M_\T^{}) = (0.5,\,4)\,$TeV\, for both plots.
}
\label{fig5:y-a-bound}
\end{center}
\vspace*{-5mm}
\end{figure}

For $\,h\to \bbbar\,$ channel, we expect that the signal ratio is always lower than one,
because the \,Br$[h\to \bbbar ]$\, is suppressed [Fig.\ref{fig:2}(a)] and for the
associate production, the cross section is proportional to
$\,\xi^{2}_{hVV} = c_\al^2 < 1$\,.\, This is why Fig.\,\ref{fig3:R-VV}(b) shows that
the $\,h\to \bbbar\,$ signal ratio (blue curve) approaches one as $\,\al \to 0,\pi\,$,\,
and becomes zero around $\,\al = \frac{\pi}{2}\,$.\,
For $\,h\to \tautaub\,$ channel, the decay branching fraction is higher than
the SM value for $\,\alpha<\frac{\pi}{2}$\, [Fig.\ref{fig:2}(a)],
and the Higgs production via gluon fusions is enhanced only around
$\,0 < \al < 0.15\pi\,$,\, depending also on the VEV ratio $\,y\,$
[Fig.\ref{fig:2}(b)].
Hence, we find that the final $\,\tautaub\,$ signal rate
receives mild enhancement for $\,0 < \alpha < 0.15\pi$\,,\,
and approaches the SM value for $\,\al \simeq 0,\,0.15\pi,\,\pi\,$,\,
as shown in Fig.\,\ref{fig3:R-VV}(b).
The current LHC experimental errors of measuring
$\,\bbbar\,$ and $\,\tautaub\,$ channels are still too large
to give significant constraint in the theory space.
But the upcoming LHC runs with 14\,TeV collision energy and higher integrated
luminosities will better probe these two channels.

Next, we preform a global fit by including the LHC (ATLAS/CMS) data \cite{LHC2013-3,LHC2013-3b},
the Tevatron data \cite{Tevatron}, and the electroweak precision tests (Sec.\,2.2).
Our model has six independent parameters
$\,(\alpha,\,x,\,y,\,r,\,z_{t}^{},\,M_{H}^{})$, relevant for this analysis.
Note that with these we can then determine $W'$ mass from Eq.\,(\ref{eq:VV'-mass}),
and $(\T,\,\B)$ masses from Eq.\,(\ref{eq:mtbTB}) in which
$\,M_S^{}=\kappa/\!\sqrt{r} = m_t^{}/(z_t^{}\!\sqrt{r\,})\,$.\,
We derive the best fit by minimizing the following $\,\chi^{2}\,$ function,
\beqa
\chi^{2} ~=~ \sum_{ij}\,
(\hat{\mu}_{i}^{}-\hat{\mu}_{i}^{\text{exp}})
(\sigma^{2})^{-1}_{ij}
(\hat{\mu}_{j}^{}-\hat{\mu}_{j}^{\text{exp}})
\,,
\eeqa
where
$\,\hat{\mu}_{j}^{}=[\sigma\times\text{Br}]_j^{}/
   [\sigma\times\text{Br}]_j^{\text{sm}}$\,
is the Higgs signal strength for each given channel,
$\,j=\gamma\gamma,\,WW^*,\,ZZ^*,\,\bbbar,\,\tautaub\,$,\, at ATLAS, CMS and Tevatron,
or, $\,\hat{\mu}_{j}^{}\,$ denotes the electroweak precision parameters
\,$(\Shat,\,\That,\,W,\,Y)$.\,
The error matrix is,
\,$(\sigma^{2})_{ij}^{}=\sigma_{i}^{}\rho_{ij}^{}\sigma_{j}^{}$\,,\,
where $\,\sigma_i^{}$\, denotes the corresponding error and
$\,\rho_{ij}^{}\,$ is the correlation matrix.
To optimize and simplify the fits, we consider the physical requirements
that $\,x^2\ll 1\,$ for reliable perturbative expansion of parameter
space (Table\,\ref{tab2:coup-V'XY}-\ref{tab3:Hig-coup}),
and $\,\sqrt{r}=\kappa/M_S^{}=\mO(1)\,$ for natural topflavor seesaw.
So, we can fairly take the sample inputs $\,(x,\,r)=(0.2,\,1)\,$.\,
Thus, we are left with four parameters $\,(\alpha,\,y,\,z_{t}^{},\,M_{H}^{})$\,
for the global fit. From these, we derive the best fit values,
$\,(\alpha,\,y,\,z_{t}^{},\,M_{H}^{})\simeq (0.13\pi,\,0.3,\,0.06,\,650\,\GeV)$,\,
with  $\,\chi^{2}/\textrm{d.o.f} \,\simeq 14/25 < 1$\,.\,
These best fit values result in,
$\,(M_{W'}^{},\,M_\T^{})\simeq (1.3,\,4)\,$TeV,\, but their allowed
$\,1\sigma\,$ mass ranges are still large.

In Fig.\,\ref{fig5:y-a-bound}, we further perform a two parameter fit
in $\,\al -M_{W'}^{}\,$ plane, by fixing two more inputs
$\,(M_H^{},\,M_\T^{}) = (0.5,\,4)\,$TeV\, around their best fit values
(since $H^0$ and $\T$ masses only appear in the oblique precision corrections
and are not so sensitive to the fit). In Fig.\,\ref{fig5:y-a-bound}(a),
we first make the fits for ATLAS and CMS data\,\cite{LHC2013-3,LHC2013-3b},
respectively, as shown by the blue and green dashed curves at 68\%\,C.L.,
where $\,h^0\to \gaga,\,WW^*,\,ZZ^*\,$ and
$\,h^0\to \bbbar,\,\tautaub\,$ channels are included for each experiment.
We see that ATLAS data favor our model over the SM, while CMS data
are still consistent with the SM point $\,(\al,\,M_{W'}^{}) = (0,\,\infty)\,$
at $1\sigma$ level.  The best fits of ATLAS and CMS data give,
$\,(\al,\,M_{W'}^{}) = (0.15\pi,\,0.85\TeV)$\, and
$\,(\al,\,M_{W'}^{}) = (0.14\pi,\,4\TeV)$,\, respectively.
These lead to $\,y=0.47~(0.1)\,$ for ATLAS (CMS) fit.
Then, in the same plot-(a),
we present the combined fit for ATLAS and CMS data together, which is depicted
by the shaded yellow contour for $\,(\al,\,M_{W'}^{})\,$ at 68\%\,C.L.
The best fit point is, $\,(\al,\,M_{W'}^{}) = (0.16\pi,\,1.33\TeV)$,\,
as marked by the red star.

Finally, we carry out a global fit by further including Tevatron data\,\cite{Tevatron}
and electroweak precision data (Sec.\,2).
This is presented in Fig.\,\ref{fig5:y-a-bound}(b),
where the shaded (red,\,blue,\,yellow) contours impose the
$\,(1\sigma,\,2\sigma,\,3\sigma)\,$ bounds, respectively.
In this global fit, we take the same sample inputs as in Fig.\,\ref{fig5:y-a-bound}(a),
\,$(x,\,r)=(0.2,\,1)$\, and $\,(M_H^{},\,M_\T^{}) = (0.5,\,4)\,$TeV.\,
With these, we derive the best fit,
\,$(\alpha,\,M_{W'}^{})=(0.13\pi,\,1.4\TeV)$,\,
with  $\,\chi^{2}/\textrm{d.o.f} \,\simeq 15/25 < 1$\,.\,
This is marked by the black star in Fig.\,\ref{fig5:y-a-bound}(b).
We see that fitting all the current direct and indirect data clearly deviates from
the decoupling limit $\,(\al,\,M_{W'}^{}) = (0,\,\infty)$,\,
which corresponds to the SM point.
Hence, our model is favored by the existing data above $1\sigma$ level, and
will be further probed by the upcoming runs at the LHC\,(14\,TeV).

In passing, a recent interesting paper studied the Higgs fit
with extra charged vector bosons and charged scalar for different class
of models \cite{Alanne:2013dra}.


\vspace*{2mm}
\section{Heavier Higgs Boson Signals at the LHC}

In this section we study the LHC signals of the heavier Higgs boson $\,H^0$,\,
which is an indispensable prediction of our Higgs sector beyond the conventional SM.
We also analyze the existing searches on a heavier Higgs boson at the
LHC\,(7\,TeV+8\,TeV), and derive constraints on the $\,H^0\,$ mass $\,M_H^{}\,$
and the Higgs mixing angle $\,\al\,$.

\begin{figure}[t]
\begin{center}
\vspace*{-1mm}
\hspace{-5mm}
\includegraphics[width=7.8cm,height=6.5cm]{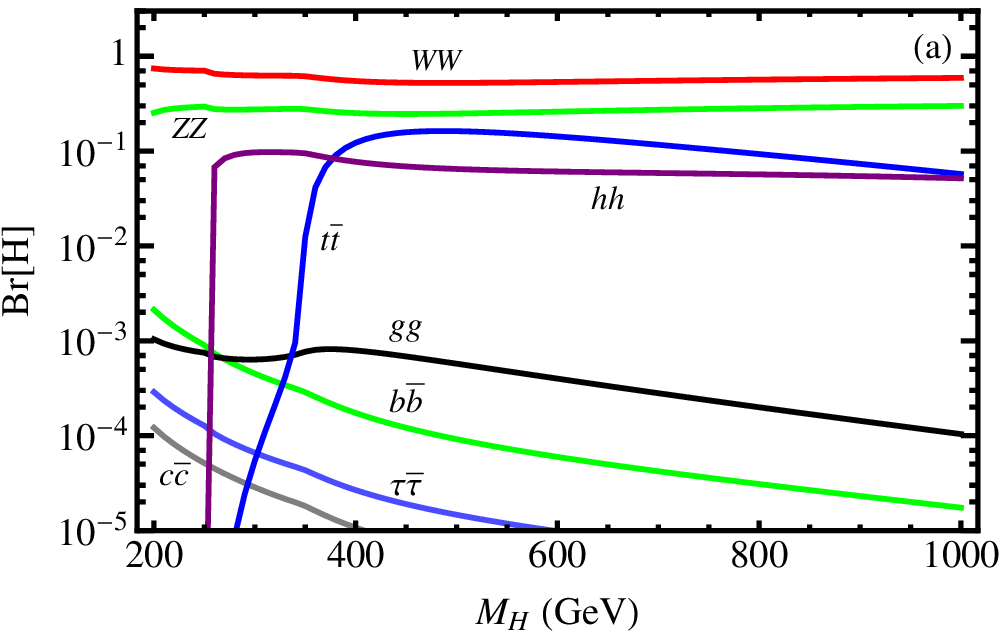}
\includegraphics[width=7.8cm,height=6.5cm]{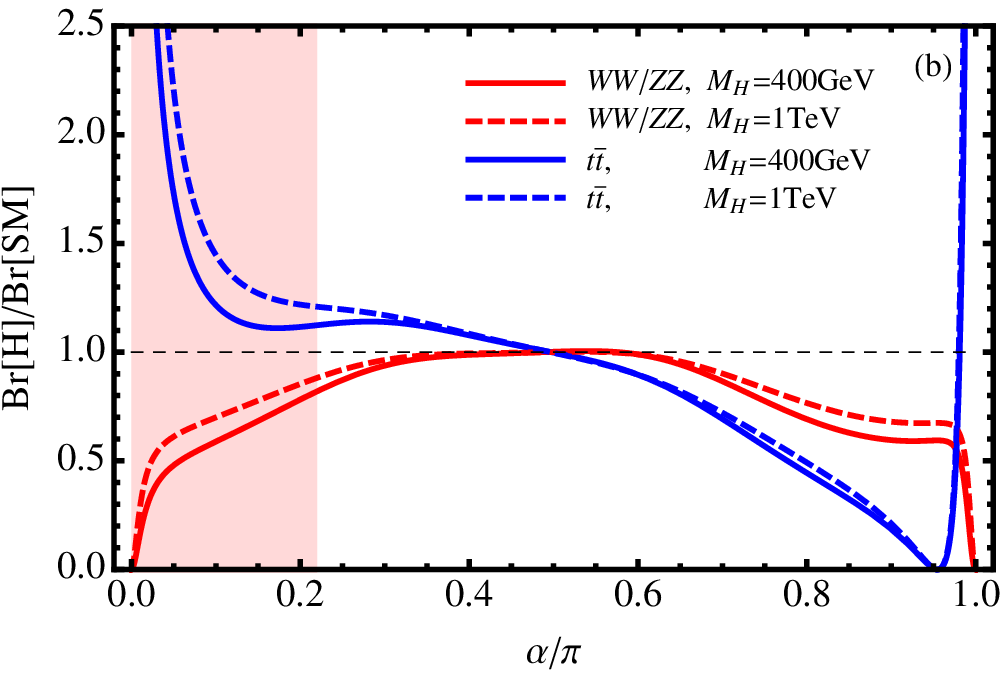}
\vspace*{-3mm}
\caption{Plot-(a) presents
decay branching fractions of heavier Higgs boson $H^0$ as functions
of the Higgs mass $\,M_H^{}\,$,\, where we input
$\,\alpha = 0.13\pi$\, based on our best fit.
Plot-(b) depicts ratios of $H^0$ decay branching fractions over the
SM values as functions of mixing angle $\,\al\,$,\,
for the three major channels $\,H^0\to WW,\,ZZ,\,t\bar{t}\,$.\,
The shaded pink band depicts the $1\sigma$ favored range
of $\,\alpha$\,,\, from our global fit in Fig.\,\ref{fig5:y-a-bound}(b).
In both plots, we have sample inputs,
\,$(y,\,r)=(0.29,\,1)$\, and $\,(M_{W'}^{},\,M_\T^{}) = (1.4,\,4)\,$TeV.\,
}
\label{fig6:H-Br}
\end{center}
\end{figure}

The gauge and Yukawa couplings of $\,H^0$\, are presented in Table\,\ref{tab3:Hig-coup}.
With these we compute the decay branching fractions of $\,H^0$\, and summarize them
in Fig.\,\ref{fig6:H-Br}(a).
We see that $\,H^0\to ZZ\,$ and $\,H^0\to WW\,$
are the two dominant decay channels at the LHC.
The other two channels $\,H^0\to t\bar{t}\,$ and $\,H^0\to hh\,$
have branching ratios generally below about 20\% and 10\%, respectively.
For comparison, we also evaluate ratios of the three major branching fractions of $\,H^0$\,
over that of a hypothetical SM Higgs boson with the same mass. These are shown
in Fig.\,\ref{fig6:H-Br}(b) as functions of the Higgs mixing angle $\,\al\,$,
for two representative Higgs masses,
$\,M_H^{}=400\,$GeV (solid curves) and $\,M_H^{}=1\,$TeV (dashed curves).\,
We see that the $\,H^0$\, decay branching fractions are rather insensitive to the Higgs
mass in the $\,WW/ZZ\,$ channels (over full $M_H^{}$ range) and $\,t\bar{t}\,$ channel
(for $\,M_H^{}> 350$\,GeV).
Furthermore, Fig.\,\ref{fig6:H-Br}(b) shows that around the best fit
range of $\,\al\,$ (marked by the pink band), the branching fractions of
$\,WW/ZZ\,$ channels are significantly lower than the SM, while the $\,t\bar{t}\,$
mode has higher branching ratio above the SM value.

In parallel to Eq.\,(\ref{eq:RggF-h}), we can define the ratio of production
cross sections,
$\,\mR_{ggF}^{}[H] = {\sigma [{gg\to H}]}/{\,\sigma[{gg\to H}]_{\rm{SM}}^{}\,}\,$.\,
Different from (\ref{eq:RggF-app}), with Table\,\ref{tab3:Hig-coup}
we can estimate the production rate of $\,H^0\,$ as follows,
\beqa
\mR_{ggF}^{}[H]
~\simeq~ \(\xi_{Htt}^{} + \xi_{H\mT\mT}^{} + \xi_{H\mB\mB}^{}\)^{2}
\,\simeq~ \(-\sa+y\frac{\ca}{\,1\!+\!r\,} - z^{2}_{t}\frac{\omega_H^{}}{\,1\!+\!r\,}\)^{2}
\,\simeq~ \(\sa - \frac{y\ca}{\,1\!+\!r\,}\)^2  \,,
\label{eq:RggF-Happ}
\eeqa
where $\,z_t^2 \ll y \ll 1 \,$.\,
We see that for small Higgs mixing angle $\,\al\,$,\,
the $\,H^0$\, production rate is much more suppressed than
that of $\,h^0\,$ in Eq.\,(\ref{eq:RggF-app}).

Next, we analyze the LHC constraints and potentials for probing
the heavier Higgs boson $\,H^0$.\,
The latest ATLAS and CMS data \cite{LHC2013-3,LHC2013-3b} have excluded the mass of
a SM Higgs boson up to 650\,GeV and 800\,GeV at 95\% confidence level, respectively.
The most sensitive detection channels for a heavier SM-like Higgs boson are
$\,H\to WW,ZZ\,$.\, The major production mechanisms for a heavier SM-like Higgs boson
are the gluon fusions and vector boson fusions. At the LHC(8\,TeV) and LHC(14\,TeV),
the gluon fusions always give the largest production cross section over the
wide Higgs-mass range up to about 1\,TeV \cite{LHC-CX}.
Hence, we will focus on
$\,pp\to H^0\to ZZ\to 4\ell\,$ and
$\,pp\to H^0\to WW\to 2\ell 2\nu\,$
processes for detecting $H^0$ at the LHC.

\begin{figure}
\begin{center}
\vspace*{-3mm}
\hspace{-5mm}
\includegraphics[width=8.3cm,height=7cm]{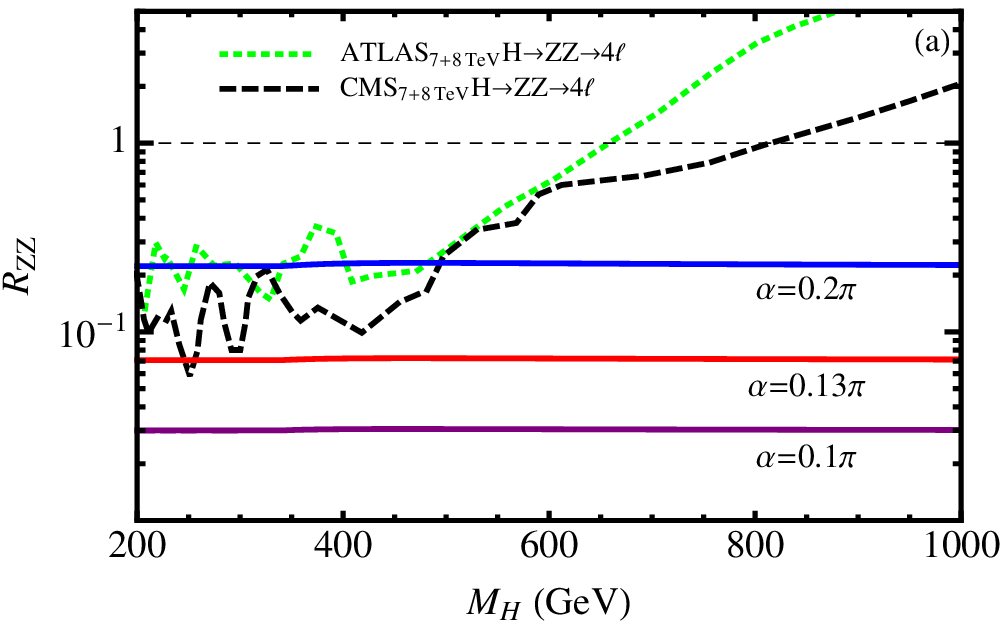}
\includegraphics[width=8.3cm,height=7cm]{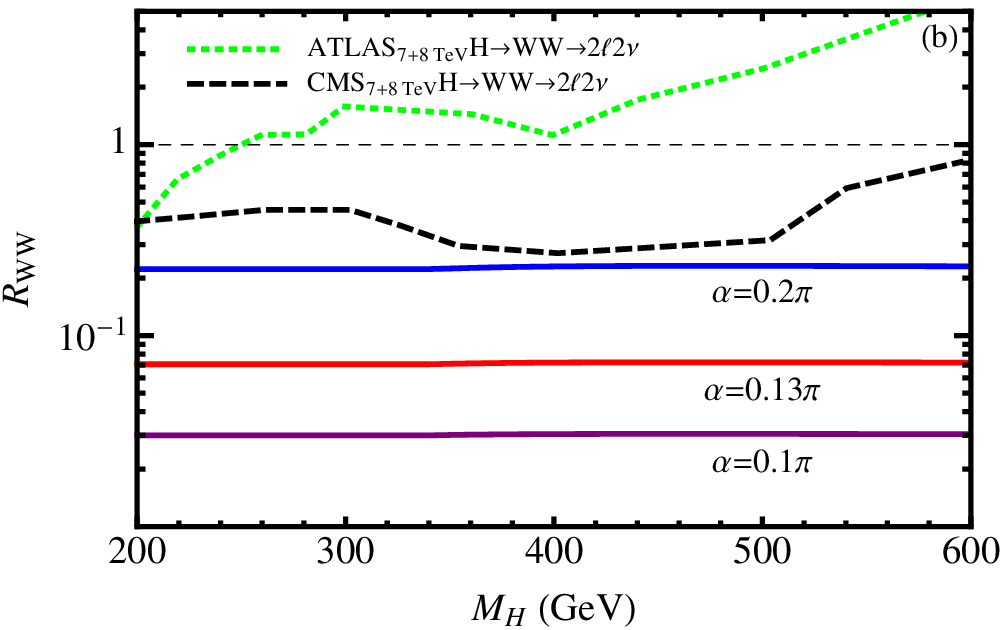}
\hspace{-5mm}
\vspace*{-3mm}
\caption{Signal rates of heavier Higgs boson $\,H^0$\, in the $\,ZZ\,$ channel
[plot-(a)] and $\,WW$\, channel [plot-(b)].
The (purple, red, blue) solid curves give our theory predictions for
$\,\al = (0.1\pi,\,0.13\pi,\,0.2\pi )$,\, respectively.
The dashed green (black) curves present the ATLAS (CMS) 95\%\,C.L.\ upper limits.
Both plots have the sample inputs
\,$(y,\,r)=(0.29,\,1)$\, and $\,(M_{W'}^{},\,M_\T^{}) = (1.4,\,4)\,$TeV,\,
based on our global fit in Fig.\,\ref{fig5:y-a-bound}(b).
}
\label{fig7:H-ZZ/WW}
\end{center}
\vspace*{-4mm}
\end{figure}

Let us define the signal rates of $\,H^0$\, over
that of a hypothetical SM Higgs boson with the same mass,
\beqs
\beqa
\mR_{ZZ}^{}[H] &=&
\frac{\,\sigma(gg\!\to\!H)\!\times\!\textrm{Br}(H\!\to\!ZZ\!\to\!4\ell)\,}
  {~[\sigma(gg\!\to\!h)\!\times\!\textrm{Br}(h\!\to\!ZZ\!\to\!4\ell)]_{\rm SM}^{}~}\,,
\\[2mm]
\mR_{WW}^{}[H] &=&
\frac{\,\sigma(gg\!\to\!H)\!\times\!\textrm{Br}(H\!\to\!WW\!\to\!2\ell 2\nu)\,}
  {~[\sigma(gg\!\to\!h)\!\times\!\textrm{Br}(h\!\to\!WW\!\to\!2\ell 2\nu)]_{\rm SM}^{}~}\,.
~~~~~~
\eeqa
\eeqs
In Fig.\,\ref{fig7:H-ZZ/WW}, we present the signal rates
$\,\mR_{ZZ}^{}[H]\,$ and $\,\mR_{WW}^{}[H]\,$ in plots (a) and (b), respectively.
For this analysis, we take the sample inputs
$\,(y,\,r,\,M_{W'}^{},\,M_{\T}^{})=(0.29,\,1,\,1.4\,\TeV,\,4\,\TeV)$\,
based on our best fits. In each plot, we also derive the predicted signal rates
for three representative Higgs mixing angles,
$\,\al = (0.1\pi,\,0.13\pi,\,0.2\pi )$,\, where $\,\alpha =0.13\pi\,$
is the best fit value. This covers significant viable range of the $\,\al\,$ angle
[cf.\ Fig.\,\ref{fig5:y-a-bound}(b)].

Figs.\,\ref{fig7:H-ZZ/WW}(a)-(b) show that the $\,ZZ$\, channel always gives
stronger bound than the $\,WW$\, channel, for both ATLAS and CMS data.
From Fig.\,\ref{fig7:H-ZZ/WW}, we find the constraints to be significantly
relaxed for smaller $\,\alpha$\, values,
such as $\,\al = 0.1\pi\,$ (purple curve) or any $\,\al < 0.13\pi\,$,
which is fully free from the current
search limits at the LHC\,(7+8\,TeV).
For the best fit  $\,\al = 0.13\pi\,$ (red curve), $H^0$ receives
almost no bound yet, except for the tiny regions around
$\,M_{H}^{} = 250\,$GeV and 300\,GeV in Fig.\,\ref{fig7:H-ZZ/WW}(a),
from the CMS searches via $ZZ$ channel.
Taking a larger value of $\,\al = 0.2\pi\,$,\,
we infer a stronger 95\%\,C.L.\ mass limit, $\,M_{H}^{} > 500\,$GeV,
from the same plot-(a).
This situation is because the cubic $\,HWW$\, and $\,HZZ$\, couplings
are proportional to $\,\sin\al\,$ (Table\,\ref{tab3:Hig-coup}),
so they become more suppressed for smaller $\,\al\,$ mixing.
It is expected that analyzing the complete data sets
of LHC\,(8\,TeV) should either place tighter bounds or reveal exciting new evidence
of such a non-SM heavier Higgs boson $H^0$.\,
The upcoming runs at the LHC(14\,TeV) will further
probe our predicted $\,H^0\,$ signals over its full mass range.


\vspace*{3mm}
\section{Conclusions}
\vspace*{2mm}

The LHC discovery of a 125\,GeV Higgs-like boson \cite{LHC2012,LHC2013-3,LHC2013-3b}
has opened up a new era for studying Higgs physics and mass generation.
It is highly anticipated that the upcoming LHC runs at 14\,TeV
with increased luminosities will further probe new physics
with the electroweak symmetry breaking and origin of masses.
The topflavor seesaw mechanism \cite{He:1999vp} provides a truly simple and
elegant renormalizable realization of the electroweak symmetry breaking
and top-mass generation, in which the top sector is special by joining
a new $SU(2)$ gauge force. It gives distinctive predictions of
vector-like spectator quarks $\,(\T,\,\B)$\,,\, new gauge bosons
\,($W',\,Z'$)\,,\, and extra heavier Higgs state \,$H^0$.\,

\vspace*{1mm}

In this Letter, we studied the LHC phenomenology of
topflavor seesaw mechanism \cite{He:1999vp}.
In Sec.\,2.1, we analyzed the structure of topflavor seesaw
including its Higgs, gauge and top sectors. We identified proper
expansion parameters and derived the mass-spectra of Higgs bosons,
gauge bosons, top/bottom quarks and spectator quarks,
as well as the associated mixing angles.
With these we presented the gauge and Yukawa couplings of Higgs bosons $\,(h^0,\,H^0)\,$
in Table\,\ref{tab3:Hig-coup}, and the $\,(W',\,Z')\,$ couplings
in Table\,\ref{tab2:coup-V'XY}.
Then, in Sec.\,2.2, we analyzed the indirect precision constraints on the theory space
(Fig.\,\ref{fig:STfit}), which push the masses of $\,(\T,\,\B)\,$ to be above $1.5-2$\,TeV,
and $(W',\,Z')$ masses above \,$0.45-1$\,TeV,\, depending on the Higgs mixing angle $\,\al\,$.\,
We further derived the LHC direct search limits on the $\,W'/Z'\,$ masses in our model.
We found, $\,M_{W'}^{}>1.25-1.6\,$TeV and $\,M_{Z'}^{}>1.0\,$TeV
at 95\%\,C.L., from the ATLAS and CMS data \cite{V'-ATLAS,V'-CMS}.

\vspace*{1mm}

In Sec.\,3, we presented analysis for decays and productions of the lighter
Higgs boson $\,h^0\,$(125\,GeV), as shown in Fig.\,\ref{fig:2}.
We derived new predictions for $h^0$ signal rates
in the $\,h^0\to\gaga,\,WW^*,ZZ^*$\, channels (via gluon fusions),
the $\,h^0\to\tautaub\,$ channel (via vector boson fusions), and
the $\,h^0\to \bbbar\,$ channel (via $Vh$ associate productions).
These are depicted in Figs.\,\ref{fig3:R-VV}(a)-(b),
where the latest ATLAS and CMS measurements\,\cite{LHC2013-3} in each channel are
displayed for comparison. We reveal that this model has significant viable
parameter space where the Higgs diphoton rate is properly enhanced, and
$\,WW^*/ZZ^*$\, and $\,\tautaub /\bbbar\,$ rates only mildly deviate from the SM.
Then, we performed the global fit by including both the direct
searches (LHC\,\cite{LHC2013-3,LHC2013-3b} and Tevatron\,\cite{Tevatron}) and the
indirect precision constraints (Sec.\,2.2).
With the proper sample inputs $\,(x,\,r)=(0.2,\,1)\,$,\, we derive the best fit,
$\,(\alpha,\,y,\,z_{t}^{},\,M_{H}^{})\simeq (0.13\pi,\,0.3,\,0.06,\,650\,\GeV)$,\,
with  $\,\chi^{2}/\textrm{d.o.f} \,\simeq 14/25 < 1$\,.\,
This also leads to, $\,(M_{W'}^{},\,M_\T^{})\simeq (1.3,\,4)\,$TeV,\, but still with
large $\,1\sigma$\, ranges.
In Fig.\,\ref{fig5:y-a-bound}, we further carried out a two-parameter $\chi^2$ fit
for $\,(\al,\,M_{W'}^{})\,$,\, where we set $H^0$ and $\T$ masses around their best fits,
$\,(M_{H}^{},\,M_\T^{})\simeq (0.5,\,4)\,$TeV.\,
Fig.\,\ref{fig5:y-a-bound}(a) presented the $\,\chi^2$ fit of $\,(\al,\,M_{W'}^{})\,$
after including all search channels at ATLAS and CMS, with the combined limit
given by the yellow contour.
In Fig.\,\ref{fig5:y-a-bound}(b), we further performed a global fit of
$\,(\al,\, M_{W'}^{})\,$  by adding the Tevatron searches and precision
constraints, which results in the best fit,
$\,(\al,\,M_{W'}^{}) = (0.13\pi,\,1.4\,\TeV)$\,,\,
with $\,\chi^{2}/\textrm{d.o.f} \,\simeq 15/25 < 1$\,.\,
Fig.\,\ref{fig5:y-a-bound}(b) also presented the \,$(1\sigma,\,2\sigma,\,3\sigma)$\,
bounds on the theory space. It shows that the current data already starts to
discriminate our model from the SM beyond $\,1\sigma$\, level.

\vspace*{1mm}

Finally, in Sec.\,4, we studied the LHC signatures of the heavier Higgs state
$\,H^0$,\, which is an indispensable prediction of our Higgs sector.
We analyzed the $\,H^0$\, decays in Fig.\,\ref{fig6:H-Br}(a)-(b), which are dominated
by the two major channels of $\,H^0\to WW,\,ZZ$.\, Their decay branching fractions
are sensitive to varying the mixing angle $\,\al\,$,\,
but remain largely unaltered over the full range of $\,H^0$\, mass.
The most sensitive detection modes come from leptonic decay products of
$\,H^0\to ZZ\to 4\ell\,$ and $\,H^0\to WW\to 2\ell 2\nu\,$.\,
In Fig.\,\ref{fig7:H-ZZ/WW}(a)-(b), we presented our new predictions of the $\,H^0$\,
signal rates via $ZZ$ and $WW$ channels,
for three sample inputs of mixing angle $\,\al = (0.1\pi,\,0.13\pi,\,0.2\pi )$,\,
consistent with our global fit in Fig.\,\ref{fig5:y-a-bound}(b).
We imposed the current LHC search limits on the theory space,
as the green and black dashed curves in each plot.
We found that for our best fit $\,\al = 0.13\pi\,$,\,
the $\,H^0$\, boson only receives a mild lower mass bound,
$\,M_{H}^{} > 250-300\,$GeV.\,
But, for $\,\al < 0.13\pi\,$,\, the $H^0$ state is fully free from
the existing LHC constraints so far.
The upcoming runs at the LHC\,(14\,TeV) with higher
integrated luminosities will have high potential to discover or exclude
the $\,H^0$\, Higgs boson through its full mass range.

\vspace*{5mm}
\noindent
{\bf Acknowledgments}
\\[1.5mm]
We thank Tomohiro Abe and Ning Chen for related discussions.
We also thank Bogdan Dobrescu and Chris Hill for discussing the topseesaw.
This work was supported by National NSF of China
(under grants 11275101, 11135003)
and National Basic Research Program (under grant 2010CB833000).


\end{document}